\documentclass[hidelinks,onefignum,onetabnum, dvipsnames]{siamart220329}

\usepackage[utf8]{inputenc} 
\usepackage[T1]{fontenc}    
\usepackage{hyperref}       
\usepackage{url}            
\usepackage{booktabs}       
\usepackage{amsfonts}       
\usepackage{amsmath}
\usepackage{nicefrac}       
\usepackage{microtype}      

\usepackage{multicol} 

\usepackage{sgame}
\usepackage{upgreek}
\usepackage{bm}
\usepackage{amsmath,amssymb,amsfonts}

\usepackage{url}

\usepackage{tikz}
\usetikzlibrary{decorations.pathreplacing}

\usepackage{caption}
\usepackage{subcaption}

\usepackage[algo2e, linesnumbered, ruled,noend]{algorithm2e}		
\SetAlFnt{\small}  %

\newcommand{\blue}[1]{{\color{black}#1}}  

\newcommand{\nn}{\mathbb{N}}

\newcommand{\rr}{\mathbb{R}}

\newcommand{\xx}{\mathbb{X}}

\newcommand{\zz}{\mathbb{Z}}

\newcommand{\NN}{\mathcal{N}}
\newcommand{\GG}{\mathcal{G}}
\newcommand{\PP}{\mathcal{P}}
\newcommand{\QQ}{\mathcal{Q}}

\newcommand{\VV}{\mathcal{V}}

\renewcommand{\P}{{\rm Pr}}

\newcommand{\bpi}{\bm{\pi}}
\newcommand{\bPi}{\bm{\Pi}}

\newcommand{\bQ}{\textbf{Q}}
\newcommand{\bH}{\mathbf{H}}
\newcommand{\bh}{\mathbf{h}}

\newcommand{\ba}{\mathbf{a}}

\newcommand{\bphi}{\bm{\phi}}
\newcommand{\balpha}{\bm{\alpha}}
\newcommand{\bomega}{\bm{\omega}}
\newcommand{\bvarpi}{\bm{\varpi}}

\newcommand{\uniform}{\text{Unif}}

\newcommand{\eq}[1][0]{\bm{\Pi}^{#1{\rm \text{-}eq}}   }

\newcommand{\BR}{{\rm BR}}

\newcommand{\simplex}{\mathcal{P}} 
\newcommand{\discount}{\gamma} 

\newcommand{\IndividualPolicySet}{\Pi} 

\newcommand{\JointPolicySet}{\bPi} 

\newcommand{\IndividualActionSet}{\mathbb{A}}     

\newcommand{\JointActionSet}{\mathbf{A}}     

\newtheorem{assumption}{Assumption}

\begin{document}

\title{Unsynchronized Decentralized Q-Learning: {Two Timescale Analysis By Persistence}\thanks{{An earlier version of this work containing preliminary results was presented at the 2023 IEEE Conference on Decision and Control.}
}
}

\author{Bora Yongacoglu\thanks{Department of Electrical and Computer Engineering, University of Toronto}
\and G\"{u}rdal Arslan\thanks{Department of Electrical Engineering, University of Hawaii at Manoa}
\and Serdar Y\"{u}ksel\thanks{Department of Mathematics and Statistics, Queen's University}
}

\maketitle

\begin{abstract}%
Non-stationarity is a fundamental challenge in multi-agent reinforcement learning (MARL), where agents update their behaviour as they learn. Many theoretical advances in MARL avoid the challenge of non-stationarity by coordinating the policy updates of agents in various ways, including synchronizing times at which agents are allowed to revise their policies. Synchronization enables analysis of many MARL algorithms via multi-timescale methods, but such synchronization is infeasible in many decentralized applications. In this paper, we study an unsynchronized variant of the decentralized Q-learning algorithm, a recent MARL algorithm for stochastic games. We provide sufficient conditions under which the unsynchronized algorithm drives play to equilibrium with high probability. Our solution utilizes constant learning rates in the Q-factor update, which we show to be critical for relaxing the synchronization assumptions of earlier work. Our analysis also applies to unsynchronized generalizations of a number of other algorithms from the regret testing tradition, whose performance is analyzed by multi-timescale methods that study Markov chains obtained via policy update dynamics. This work extends the applicability of the decentralized Q-learning algorithm and its relatives to settings in which parameters are selected in an independent manner, and tames non-stationarity without imposing the coordination assumptions of prior work.
\end{abstract}



\author{Bora Yongacoglu \thanks{Department of Mathematics and Statistics, Queen's University}
\and G\"{u}rdal Arslan \thanks{Department of Electrical Engineering, University of Hawaii at Manoa}
\and Serdar Y\"{u}ksel\footnotemark[1]
}

\thispagestyle{empty}
\pagestyle{empty}

\section{Introduction}

Multi-agent systems are characterized by the coexistence of several autonomous agents acting in a shared environment. In multi-agent reinforcement learning (MARL), agents in the system change their behaviour in response to feedback information received in previous interactions, and the system is non-stationary from any one agent's perspective. In this non-stationary environment, agents attempt to optimize their performance against a moving target \cite{hernandez2019survey}. This non-stationarity has been identified as one of the fundamental technical challenges in MARL \cite{hernandez2017survey}. In contrast to the rich literature on single-agent learning theory, the theory of MARL is relatively underdeveloped, due in large part to the inherent challenges of non-stationarity and decentralized information.

This paper considers learning algorithms for stochastic games, a common framework for studying MARL in which the cost-relevant history of the system is summarized by a state variable.  We focus on stochastic games in which each agent  fully observes the system's state variable but does not observe the actions of other agents, exacerbating the challenge of non-stationarity.

Early theoretical work on MARL in stochastic games avoided the problem of non-stationarity by studying applications in which joint actions were observed by all agents \cite{littman1996generalized, Littman2001ffq, hu2003nash}. There has also been interest in the \emph{local action learner} setting, where actions are not shared between agents. Several rigorous contributions have recently been made in this setting, including the works of \cite{arslan2017decentralized,daskalakis2020independent, sayin2021decentralized, yongacoglu2022decentralized, yongacoglu2023satisficing}, to be discussed shortly.

Of the recent advances in MARL for the local action learner setting, many of the algorithms with strong guarantees have circumvented the challenge of non-stationarity by relying, implicitly or explicitly, on coordination between agents. In particular, several works in the \emph{regret testing} tradition rely on some form of synchronization, whereby agents agree on the times at which they may revise their behaviour and are constrained to fix their policies during the intervening periods. This synchronized approach allows for the use of two timescale analyses common to many single- and multi-agent reinforcement learning paradigms, in which the behaviour of value function estimates and the behaviour of the joint policy process can be effectively uncoupled and analyzed separately  \cite{leslie2003convergent,borkar2002reinforcement,konda1999actor}.

A variety of regret testing algorithms have been proposed for different classes of games, including \cite{foster2006regret,germano2007global,arslan2017decentralized} and \cite{yongacoglu2023satisficing}. These algorithms differ in their policy update rules, which are typically chosen to exploit some underlying structure of particular classes of games, from better/best-response dynamics in \cite{arslan2017decentralized} to arbitrary $\epsilon$-satisficing policy update rules in \cite{yongacoglu2023satisficing}. Despite such differences, previous regret testing algorithms are united in their reliance on synchronization to facilitate the analysis of learning iterates and to guarantee accuracy of value function estimates. \blue{While this is justifiable in some settings, it can be restrictive in others, including applications where parameters are selected in a distributed or decentralized manner, as well as settings in which parameter coordination can be affected by communication noise.} As such, it would be desirable to provide MARL algorithms that do not require synchronization but still come with rigorous performance guarantees in the local action learner setting.

This paper studies weakly acyclic stochastic games, where the definition of weak acyclicity is in terms of single-agent best-responding. In this context, our aim is to give guarantees of convergence to stationary deterministic equilibrium policies. We find that, with slight modification, the decentralized Q-learning algorithm of  \cite{arslan2017decentralized} can be made to tolerate unsynchronized policy updates, and that strong coordination assumptions about parameter choice and timing are not necessary. This constitutes a direct generalization of the algorithm and results in \cite{arslan2017decentralized}.  The key modification involves utilizing constant learning rates in the Q-function updates, replacing the decreasing learning rates used in \cite{arslan2017decentralized} and in all previous regret testing algorithms. Unlike decreasing learning rates, which place relatively large weight on data encountered in the distant past, constant learning rates allow agents to rapidly correct their Q-factor estimates by discarding outdated information. Additionally, we retain the use of random inertia in the policy update, which we show acts as a decentralized stabilizing mechanism. 

Although the technical exposition here focuses on weakly acyclic games, the analytic insights in this paper are not limited to this setting. In fact, the results of this paper can be generalized in various directions by studying different classes of games, different policy revision mechanisms, or by selecting different sets of policies as the target for convergence. We now list several other regret testing results that can be generalized to accommodate for unsynchronized agents by employing constant learning rates and inertia in the policy update.

\textit{Teams and Optimality.} Stochastic teams and their generalizations of common interest games constitute an important special case of weakly acyclic games, with applications to distributed control \cite{yongacoglu2022decentralized}. In common interest games, one typically seeks \emph{team optimal policies}, a subset of equilibrium policies that simultaneously achieves minimum cost for each player and may not exist in more general classes of games. A modification of the decentralized Q-learning algorithm was proposed in  \cite{yongacoglu2022decentralized}, which incorporated memory to account for past performance as well as random search of players' policy spaces to search for team optimal policies. This algorithm, however, was developed in the synchronized setting, where the analysis of learning iterates was decoupled from policy dynamics. This algorithm can be further adapted to accommodate unsynchronized policy updates by utilizing constant learning rates.

\textit{General games and $\epsilon$-equilibrium.} Rather than restricting players to revise policies via inertial best-responding, one may be interested in more general revision mechanisms, to model settings where players are more exploratory in nature. If one imposes a stopping condition, such as refraining from updating when one is already $\epsilon$-best-responding, then the policy dynamics can be studied using the theory of \emph{$\epsilon$-satisficing paths} presented by \cite{yongacoglu2023satisficing}. In \cite{yongacoglu2023satisficing}, it was shown that a synchronized variant of decentralized Q-learning, which used random search of the policy space when not $\epsilon$-best-responding, drove policies to $\epsilon$-equilibrium in symmetric stochastic games. With appropriate modifications, a similar result can be obtained without synchronization.

\textit{General games and ``cumber" sets.} In general stochastic games, the set of stationary deterministic equilibrium policies can be empty. Even if this set is not empty, in general, it is not guaranteed that inertial best-response dynamics drives play to equilibrium. The notions of \emph{cusber} sets (introduced in \cite{josephson2004stochastic}) and \emph{cumber} sets (\cite{yongacoglu2022decentralized}) are useful for studying the dynamics of inertial best-responding in general games. Cusber sets are subsets of policies that are closed under single-agent best-responses, while cumber sets are defined as being closed under multi-agent best-responses. It was previously shown in \cite{yongacoglu2022decentralized} that variants of decentralized Q-learning drive play to minimal cumber sets even in general games, though play may cycle between several policies within this minimal set. The results of the present paper can be combined with the analysis of \cite{yongacoglu2022decentralized} to guarantee convergence of unsynchronized decentralized Q-learning to cumber sets in general games.

\vspace{7.5pt} 

\noindent \textbf{Contributions:} 
In this paper, we offer a means to relax the synchronization assumptions common in the regret testing paradigm of MARL while retaining desirable convergence guarantees. Our solution combines inertial policy updating with persistent learning via non-decreasing learning rates. 
We provide mathematical details for a particular set-up involving the unsynchronized variant of the decentralized Q-learning algorithm of \cite{arslan2017decentralized}, a recent MARL algorithm designed for weakly acyclic stochastic games, a class of games that admits teams and potential games as important special cases. We do not impose synchronization of policy revision times, which is a common and restrictive assumption in prior work. We show that persistent learning, when combined with inertial policy updating, provides a mechanism for taming non-stationarity and mitigating the moving target problem of MARL \emph{without} coordinating agents' parameter choices ahead of play. In particular, we show that this algorithm drives policies to equilibrium with arbitrarily high probability, under appropriate parameter selection. As we discuss below in \Cref{sec:literature}, this appears to be the first formal analysis of an unsynchronized algorithm in the \emph{regret testing} tradition. We also note that the same analysis can be adapted and applied to a variety of related learning algorithms, which all impose the synchronization condition and are analyzed via multi-timescale methods. Thus, the results of this paper establish that persistent learning can address analytic challenges relating to unsynchronized algorithms in a variety of multi-agent learning environments above and beyond weakly acyclic games.

\vspace{5pt}

\noindent \textbf{Notation:} For a standard Borel space $A$, we let $\simplex ( A )$ denote the set of probability measures on $A$ with its Borel $\sigma$-algebra. For standard Borel spaces $A$ and $B$, we let $\PP ( A | B )$ denote the set of transition kernels on $A$ given $B$. For a finite set $A$, we let $\rr^A$ denote the set of functions from $A$ to $\rr$. We let $\uniform( A )$ denote the uniform probability distribution on a compact set $A$. We let $\zz_{\geq 0}$ denote the set of non-negative integers. To denote that a random variable $X$ has distribution $\mu$, we write $X \sim \mu$. 

\vspace{2.5pt}

\textbf{Terminology:} We use the terms \emph{player} and \emph{agent} interchangeably. When considering a particular player and referring to other players in the same game, these other players may be variously called \emph{counterparts} or \emph{counterplayers}.

\vspace{2.5pt}

\subsection{Organization} The remainder of the paper is organized as follows. In \Cref{sec:model}, we present the model of stochastic games. Related work is surveyed in \Cref{sec:literature}. In \Cref{sec:learning-algorithms}, we present the unsynchronized decentralized Q-learning algorithm and we state our main result, \Cref{theorem:main}, on high probability guarantees for equilibrium in weakly acyclic stochastic games. \blue{The proof of \Cref{theorem:main} is outlined in \Cref{sec:proof-outline}.}   The details of the proof of \Cref{theorem:main} are presented in \Cref{sec:proof}. The results of a simulation study are presented in \Cref{sec:simulations}, and the final section concludes. 
\blue{A glossary of notation is provided in Appendix~\ref{appendix:notation}.}

\section{Model} \label{sec:model} 

A stochastic game $\GG$ is described by a list:  
\begin{equation} \label{eq:stochastic-game} 
\GG = \left( \NN,    \xx,  \{ \IndividualActionSet^i, c^i, \discount^i : i \in \NN \} , P , \nu_0 \right). 
\end{equation}

\noindent The components of $\GG$ are as follows: $\NN = \{ 1, 2, \dots, N \}$ is a finite set of $N$ agents. The set $\xx$ is a finite set of system states. For each player $i \in \NN$, $\IndividualActionSet^i$ is $i$'s finite set of actions. We write $\JointActionSet = \times_{i \in \NN} \IndividualActionSet^i$, and refer to elements of $\JointActionSet$ as \emph{joint actions}. For each player $i$, a function $c^i : \xx \times \JointActionSet \to \rr$ determines player $i$'s stage costs, which are aggregated using a discount factor $\discount^i \in [0,1)$. The initial system state has distribution $\nu_0 \in \simplex ( \xx )$, and state transitions are governed by a transition kernel $P  \in \PP ( \xx | \xx \times \JointActionSet )$.

At time $t \in \zz_{\geq 0}$, the state variable is denoted by $x_t$, and each player $i$ selects an action $a^i_t \in \IndividualActionSet^i$ according to its policy, to be described shortly. The joint action at time $t$ is denoted $\ba_t$. Each player $i$ then incurs a cost $c^i_t := c^i ( x_t, \ba_t )$, and state variable evolves according to $x_{t+1} \sim P  ( \cdot | x_t, \ba_t )$. This process is then repeated at time $t+1$, and so on. 

A policy for player $i$ is an action selection rule that, at each time $t \geq 0$, selects player $i$'s action $a^i_t$ in some (possibly random) manner that depends only on information locally available to player $i$ at time $t$. Using $h^i_t$ to denote the information that is locally available to player $i$ at time $t$, the following standing assumption specifies the decentralized information structure considered in this paper.

\begin{assumption}[Local action learners]  \label{ass:local-action-learners}
For each $i \in \NN$, player $i$'s information variables $\{ h^i_t \}_{t \geq 0}$ are given by $h^i_0 = x_0$ and 
\[
h^i_{t+1} = \left(	h^i_t, a^i_t, c^i ( x_t, \ba_t ) , x_{t+1} 	\right), \text{ for } t \geq 0. 
\]
\end{assumption}

Under this information structure, each player $i$ observes the complete system state without noise, and additionally observes its own actions and its own cost \emph{realizations,} but does not observe the actions of other players directly. This \emph{local action learner} (LAL) information structure is common in the literature in MARL, with examples such as \cite{arslan2017decentralized, claus1998dynamics, daskalakis2020independent, matignon2012survey, sayin2021decentralized, yongacoglu2022decentralized, yongacoglu2023satisficing}, and is sometimes called the independent learning paradigm.\footnote{This terminology is, however, not uniform, as independent learning has also recently been used to refer to learners that update their policies in a (perhaps myopic) self-interested manner \cite{ozdaglar2021independent}. In response to the changing conventions, we recommend that learners who condition only on their \textit{local} action history should be called \textit{local action learners}, in analogy to \textit{joint action learners}.} The LAL paradigm is one of the principal alternatives to the joint action learner (JAL) paradigm, studied previously in \cite{Littman1994} and \cite{Littman2001ffq} among others. The key difference is that the JAL paradigm assumes that each agent $i$ gets to observe the complete joint action profile $\ba_t$ after it is played. (That is, in the JAL paradigm, one assumes $h^i_{t+1} = ( h^i_t, \ba_t, c^i ( x_t, \ba_t), x_{t+1})$.)

\begin{definition}  \label{def:policy}
For $i \in \NN$ and $t \geq 0$, let $H^i_t := \xx \times \left( \IndividualActionSet^i \times \rr \times \xx \right)^t$. A sequence $\pi^i = ( \pi^i_t )_{t \geq 0}$ is called a \emph{policy} for player $i$ if $\pi^i_t \in \PP ( \IndividualActionSet^i | H^i_t )$ for each $t \geq 0$. 
\end{definition} 

We let $\IndividualPolicySet^i$ denote player $i$'s set of policies. When following a policy $\pi^i  = ( \pi^i_t )_{t \geq 0}$, player $i$ selects actions according to $a^i_t \sim \pi^i_t ( \cdot | h^i_t )$. 

In general, action selection can incorporate randomness, and players may use arbitrarily complicated, history-dependent policies. However, our analysis will focus on stationary (Markov) policies, a subset of policies that randomly select actions in a time invariant manner that conditions only on the currently observed state variable. Such a restriction entails no loss in optimality for a particular player, provided the remaining players also use stationary policies. Focusing on stationary policies is quite natural: we refer the reader to \cite{levy2013discounted} for an excellent elaboration. 

\begin{definition} \label{def:stationary-policy}
For $i \in \NN$, a policy $\pi^i \in \IndividualPolicySet^i$ is called \emph{stationary}, if the following holds for any $t, k \geq 0$ and any $\hat{h}^i_t \in H^i_t, \bar{h}^i_k \in H^i_k$: 
\[
\hat{x}_t = \bar{x}_k \Rightarrow \pi^i_t ( \cdot | \hat{h}^i_t ) = \pi^i_t ( \cdot | \bar{h}^i_k ),
\]
where $ \hat{x}_t$ and $ \bar{x}_k$ denote the final components of $\hat{h}^i_t$ and $\bar{h}^i_k$, respectively. 
\end{definition} 

The set of stationary policies for player $i \in \NN$ is denoted $\IndividualPolicySet^i_{S}$ and we identify $\IndividualPolicySet^i_{S}$ with $\PP ( \IndividualActionSet^i | \xx )$, the set of transition kernels on $\IndividualActionSet^i$ given $\xx$. Henceforth, unqualified reference to a policy shall be understood to mean a stationary policy, while non-stationary policies will be explicitly identified as such.

\vspace{2.5pt} 

\begin{definition} 
For $i \in \NN$, $\xi > 0$, a policy $\pi^i \in \IndividualPolicySet^i_{S}$ is called \emph{$\xi$-soft} if $\pi^i ( a^i |  x ) \geq \xi$ for all $(x,a^i)  \in \xx \times \IndividualActionSet^i$. A policy $\pi^i \in \IndividualPolicySet^i_{S}$ is called \emph{soft} if it is $\xi$-soft for some $\xi > 0$. 
\end{definition}

\vspace{1pt} 

\begin{definition} \label{def:deterministic-policy}
A policy $\pi^i \in \IndividualPolicySet^i_{S}$ is called \emph{deterministic} if for each $x \in \xx$, there exists $a^i \in \IndividualActionSet^i$ such that $\pi^i ( a^i | x ) = 1$. 
\end{definition}

\vspace{1pt} 

The set of deterministic stationary policies for player $i$ is denoted by $\IndividualPolicySet^i_{SD}$ and is identified with the set of functions from $\xx$ to $\IndividualActionSet^i$.

\vspace{5pt}

\textbf{Notation:} We let $\JointPolicySet_{S} := \times_{i \in \NN} \IndividualPolicySet^i_{S}$ denote the set of \textit{joint policies}. To isolate player $i$'s component in a particular joint policy $\bpi \in \JointPolicySet_{S}$, we write $\bpi = ( \pi^i, \bpi^{-i} )$, where $-i$ is used in the agent index to represent all agents other than $i$. Similarly, we write the joint policy set as $\JointPolicySet_{S} = \IndividualPolicySet_{S}^i \times \JointPolicySet_{S}^{-i}$, and so on.

For any joint policy $\bpi $ and initial distribution $\nu \in \simplex ( \xx )$, there is a unique probability measure on the set of state-action trajectories $( \xx \times \JointActionSet )^{\infty}$. We denote this measure by $\P^{\bpi}_{\nu}$, and let $E^{\bpi}_{\nu}$ denote its expectation. We use this to define player $i$'s value function:
\[
J^i ( \bpi,  \nu ) := 
E^{\bpi}_{\nu} \left[ \sum_{t = 0}^{\infty} (\discount^i)^t c^i ( x_t, \ba_t ) \right]. 
\]

\noindent When $\nu = \delta_{s}$ is the Dirac measure of some state $s \in \xx$, we write $J^i ( \bpi, s )$ instead of $J^i ( \bpi, \delta_{s})$. For $\bpi = ( \pi^i, \bpi^{-i})$, we will also write $J^i ( \pi^i, \bpi^{-i}, \nu ) $ to isolate the role of $\pi^i$. 

\vspace{2.5pt}
\begin{definition}  \label{def:best-response}
Let $\epsilon \geq 0$, $i \in \NN$. A policy $\pi^{*i} \in \IndividualPolicySet^i_{S}$ is called an $\epsilon$-\emph{best-response to $\bpi^{-i} \in \JointPolicySet^{-i}_{S}$} if, for every $s \in \xx$,
\[
J^i ( \pi^{*i}, \bpi^{-i}  , s )  \leq \inf_{ \tilde{\pi}^i \in \IndividualPolicySet^i } J^i ( \tilde{\pi}^i  , \bpi^{-i} ,  s ) + \epsilon .
\] 
\end{definition}

The set of $\epsilon$-best-responses to $\bpi^{-i}$ is denoted $\BR^i_{\epsilon} ( \bpi^{-i} ) \subseteq \IndividualPolicySet^i_{S}$. It is well-known that for any $\bpi^{-i} \in \JointPolicySet^{-i}_{S}$, player $i$'s set of 0-best-responses $\BR^i_0 ( \bpi^{-i} )$ is non-empty, and the infimum above is in fact attained by some policy in $\IndividualPolicySet^i_{SD}$: that is, 
$\IndividualPolicySet^i_{SD} \cap \BR^i_{0} ( \bpi^{-i} ) \not= \varnothing$. 

\vspace{2.5pt} 

\begin{definition}  \label{def:equilibrium}
Let $\epsilon \geq 0$. A joint policy $\bpi^{*} \in \JointPolicySet_{S}$ is called an \emph{$\epsilon$-equilibrium}  if $\pi^{*i} \in \BR^i_{\epsilon} ( \bpi^{*-i} )$ for all $i \in \NN$.
\end{definition}

\vspace{2.5pt} 

For $\epsilon \geq 0$, we let $\eq[\epsilon]_{S} \subseteq \JointPolicySet_{S} $ denote the set of $\epsilon$-equilibrium policies. It is known that the set $\eq[0]_{S}$ is non-empty \cite{fink1964equilibrium}. We also let $\eq[\epsilon]_{SD} \subseteq \JointPolicySet_{SD}$ denote the subset of stationary deterministic $\epsilon$-equilibrium policies, which may be empty in general.

\subsection{Weakly Acyclic Stochastic Games}

We now introduce weakly acyclic games, an important subclass of games that will be the main focus of this paper.

\begin{definition} 
A sequence $\{ \bpi_k \}_{k \geq 0}$ in $\JointPolicySet_{SD}$ is called a \emph{strict best-response path} if for any $k \geq 0$ there is a unique player $i \in \NN$ such that $\pi^{i}_{k+1} \not= \pi^{i}_k$ and $\pi^i_{k+1} \in \BR^i_0 ( \bpi^{-i}_k )$. 
\end{definition}

\vspace{2.5pt}

\begin{definition}
The stochastic game $\GG$ is weakly acyclic if \textnormal{(i)} $\eq[0]_{SD} \not= \varnothing$, and \textnormal{(ii)} for any $\bpi_0 \in \JointPolicySet_{SD}$, there is a strict best-response path from $\bpi_0$ to some $\bpi^{*} \in \eq[0]_{SD}$. 
\end{definition}

 The multi-state formulation above was stated in \cite{arslan2017decentralized}, though weakly acyclic games had previously been studied in stateless games \cite{young2004strategic}. An important special case is that of stochastic teams, where $c^i = c^j$ and $\discount^i  = \discount^j$ for any players $i, j \in \NN$, and the interests of all agents are perfectly aligned. Markov potential games with finitely many states constitute another special case of weakly acyclic games \cite{mguni2021learning, leonardos2022global,zhang2021gradient}.

\subsection{Q-Functions in Stochastic Games}

In the stochastic game $\GG$, when player $i$'s counterplayers use a stationary policy $\bpi^{-i} \in \JointPolicySet^{-i}_{S}$, player $i$ faces an environment that is equivalent to a single-agent MDP. The MDP in question depends on the policy $\bpi^{-i}$ as well as the game $\GG$, and (stationary Markov) optimal policies for this MDP are equivalent to 0-best-responses to $\bpi^{-i}$ in the game $\GG$. 

Player $i$'s best-responses to a policy $\bpi^{-i} \in \JointPolicySet^{-i}_{S}$ can be characterized using an appropriately defined \emph{Q-function}, $Q^{*i}_{\bpi^{-i}} : \xx \times \IndividualActionSet^i \to \rr$.\footnote{We use the terms Q-function and (state-) action-value function interchangeably.} The function $Q^{*i}_{\bpi^{-i}}$ can be defined by a fixed point equation of a Bellman operator, but here we give an equivalent definition in terms of the optimal policy of the corresponding MDP:
\begin{equation}
Q^{*i}_{\bpi^{-i}} ( x, a^i ) := E^{\bpi^{*}}_{\nu} \left[ \sum_{t =0}^{\infty} (\discount^i)^t c^i ( x_t, \ba_t ) \middle| x_0 = x, u^i_0 = a^i \right],      \label{eq:Q-factors}
\end{equation}
for all $(x,a^i ) \in \xx \times \IndividualActionSet^i,$ where $\bpi^{*} = ( \pi^{*i} , \bpi^{-i} )$ and $\pi^{*i} \in \BR^i_0 ( \bpi^{-i} ) \cap \IndividualPolicySet^i_{SD}. $

\vspace{2.5pt}
\begin{definition} \label{def:greedy-policy}
For $Q^i : \xx \times \IndividualActionSet^i \to \rr$ and $\epsilon \geq 0$, we define 
\[
\widehat{\BR}^i_{\epsilon} ( Q^i )  := \left\{ \pi^{*i} \in \IndividualPolicySet^i_{SD} :   Q^i ( x, \pi^{*i} ( x )) \leq \min_{a^i \in \IndividualActionSet^i } Q^i ( x, a^i ) + \epsilon ,  \: \forall x \in \xx \right\}.
\]
\end{definition}

\vspace{2.5pt}
  
The set $\widehat{\BR}^i_{\epsilon} ( Q^i )$ is the set of stationary deterministic policies that are $\epsilon$-greedy with respect to $Q^{i}$. The function $Q^i$ plays the role of an action-value function, and for $Q^i = Q^{*i}_{\bpi^{-i}}$, we have $\widehat{\BR}^i_{0} ( Q^{*i}_{\bpi^{-i}} ) = \BR^i_{0} ( \bpi^{-i}) \cap \IndividualPolicySet^i_{SD}$. 

\

When the remaining players follow a stationary policy, player $i$ can use Q-learning  to estimate its action-values, which can then be used to estimate a 0-best-response policy. In the following sections, we observe that the situation is more complicated when the remaining players employ learning algorithms that adapt to feedback and revise their action selection probabilities over time.

\section{Related Work} \label{sec:literature}

In this paper, we are interested in MARL for stochastic games in the local action learner setting of \Cref{ass:local-action-learners}. In this framework, players do not have knowledge of their cost functions, the transition kernel $P$, and perhaps even the existence or nature of other players in the system. We focus on the online setting, where players do not have access to a generative model for sampling state transitions or costs, and instead must rely on sequential interactions with the multi-agent system in order to gather data.

At a high level, our objective is to study and design learning algorithms that drive play of the game to stable and individually rational joint policies. Our objectives are closely aligned with the desiderata of MARL algorithms outlined in \cite{Bowling}, namely \emph{convergence} and \emph{rationality}. In particular, we wish to establish that joint policies converge to some stationary ($\epsilon$-) equilibrium policy.

A number of MARL algorithms have been proposed with the same high-level objectives as ours. The wide ranging field of algorithms differ from one another in terms of how and when incoming feedback data is processed and how and when the agent's policy is accordingly modified. In this section, we survey some related work that is especially relevant to our chosen setting of MARL in stochastic games with local action learners.

\subsection{Q-Learning and its Challenges}

In decentralized MARL settings, where agents have minimal information about strategic interactions with other players, one natural and commonly studied approach to algorithm design is to treat each agent's game environment as a separate single-agent RL environment.  Indeed, when the remaining players use a stationary policy $\bpi^{-i} \in \JointPolicySet^{-i}_{S}$, then player $i$ truly does face a strategic environment equivalent to an MDP, and can approach its learning task with single-agent Q-learning. 

Picking some $\widehat{Q}^i_0 \in \rr^{\xx \times \IndividualActionSet^i}$ to serve as an initial estimate of its optimal Q-function, player $i$ would iteratively apply the following update rule: 

\begin{equation}
\widehat{Q}^i_{t+1} ( x_t, a^i_t ) = ( 1 - \alpha_t ) \widehat{Q}^i_t ( x_t, a^i_t ) + \alpha_t \left[ c^i ( x_t, \ba_t ) + \discount^i \min_{ a^i } \widehat{Q}^i_t ( x_{t+1}, a^i ) \right], \label{eq:single-agent-Q-factor-update}
\end{equation} 

where $\alpha_t =  \alpha_t ( x_t, a^i_t )$ is a random (that is, history-dependent) learning rate that depends on the state-action pair $(x_t, a^i_t )$ and $\widehat{Q}^i_{t+1} ( x,a) = \widehat{Q}^i_t ( x,a)$ for all $(x,a) \not= (x_t, a^i_t )$. 

\vspace{2.5pt} 

If the learning rate $\alpha_t ( x_t, a^i_t )$ is decreased at an appropriate rate in the number of visits to the state-action pair $(x_t, a^i_t )$, and if every state-action pair is visited infinitely often, then one has $\widehat{Q}^i_t \to Q^{*i}_{\bpi^{-i}}$ almost surely \cite{tsitsiklis1994asynchronous}. We recall that $Q^{*i}_{\bpi^{-i}}$ was defined in Section~\ref{sec:model} as the optimal Q-function for player $i$ when playing against the stationary policy $\bpi^{-i}$.

In order to asymptotically respond (nearly) optimally to the policy of its counterparts, player $i$ must exploit the information carried by its learning iterates. Since one must balance this exploitation with persistent exploration, a standard approach in algorithm design is to periodically revise an agent's policy---either after every system interaction or less frequently---in a manner that places higher probability on actions with better action values. A widely used device for balancing exploration and exploitation is Boltzmann action selection, wherein
\[
\P \left( a^i_t = a^i \middle| h^i_t \right) = \frac { \exp \left(- \widehat{Q}^i_t ( x_t, {a}^i) / \tau_t \right)  } { \sum_{ \bar{a}^i \in \IndividualActionSet^i } \exp \left( - \widehat{Q}^i_t ( x_t, \bar{a}^i) / \tau_t  \right) } , \text{ for } t \geq 0 ,
\] 

where $\tau_t > 0$ is a temperature parameter that is typically taken to decrease to 0 as $t \to \infty$, and player $i$'s Q-factor estimates $\widehat{Q}^i_t$ are constructed from the local history $h^i_t$ observed by player $i$. 

Boltzmann action selection is not the only feasible alternative for balancing the trade-off between exploration and exploitation in this context; several other \textit{greedy in the limit with infinite exploration} (GLIE) mechanisms may also be used to the same effect \cite{singh2000convergence}. 

In the preceding equations, one observes that $\P ( a^i_t = \cdot \, | h^i_t )$ is simply an alternative representation of the policy actually followed by player $i$. Since player $i$'s Q-factor estimates and other relevant parameters (e.g. the temperature sequence) change with time, any adaptive learning algorithm that exploits learned iterates will be \emph{non-stationary}. If other players use analogous learning algorithms and exploit their own learned information, then other players will also follow non-stationary policies. In symbols, this is to say that $\bpi^{-i} \notin \JointPolicySet^{-i}_{S}$. In this case, the strategic environment defined by the non-stationary policy $\bpi^{-i}$ is \emph{not} generally equivalent to an MDP, and there is no guarantee that the Q-factor estimates given by \eqref{eq:single-agent-Q-factor-update} will converge to some meaningful quantity, if they are to converge at all. 

Provable non-convergence of Q-iterates obtained by single-agent Q-learning in small, stateless games has been established \cite{leslie2005individual}. Similar non-convergence had been previously been observed empirically in stochastic games \cite{Bowling, tan1993multi}. Thus, some modification is needed in order to employ algorithms based on Q-learning in multi-agent settings. Modifications may involve changes to how and when Q-factor updates are performed or may involve changes to how and when action selection probabilities are updated. 


\subsection{Algorithms for Local Action Learners}

As described in the previous section, myopic algorithms that repeatedly change an agent's policy during the course of interaction can be effective in single-agent settings, when an agent faces a stationary environment. By contrast, when all agents employ such algorithms in a multi-agent system, each agent faces a non-stationary environment and the same convergence guarantees need not apply. In addition to the previously described non-convergence of Q-factors, provable non-convergence of policies under particular learning algorithms has also been observed in various settings, \cite{foster1998nonconvergence,mazumdar2019policy,mertikopoulos2018cycles}.

With the understanding that some modification of single-agent RL algorithms is needed to address the challenge of non-stationarity in MARL, several multi-agent algorithms have been proposed that use \textit{multiple timescales}. The key feature of the multi-timescale approach is that certain iterates---whether policy parameters, Q-factor estimates, or otherwise---are updated slowly relative to other iterates. In effect, this allows one to analyze the rapidly changing iterates as though the slowly changing iterates were held fixed, artificially mimicking stationarity.

Some works have studied MARL algorithms in which a subset of agents update their policies slowly while the remaining agents update their policy parameters quickly. A two-timescale policy gradient algorithm was studied for two-player, zero-sum games in \cite{daskalakis2020independent}, where a performance guarantee was proven for the minimizing player. Other works, including \cite{nekoei2023dealing}, have empirically studied MARL algorithms in which the agents alternate between updating their policies slowly and quickly according to some commonly accepted schedule. 

A second type of multi-timescale algorithm updates various policy-relevant parameters quickly relative to value function estimates, which are updated relatively slowly. This approach can be found in \cite{ozdaglar2021independent,sayin2021decentralized} and \cite{sayin2022logit}. This can be contrasted with a third type of multi-timescale algorithm, in which policies are updated relatively slowly while value functions are updated relatively quickly. For a recent example of this, see \cite{maheshwari2022independent}.

\blue{A third type of algorithm that retains the same essential quality of multiple timescale algorithms is the regret testing approach, outlined in the next subsection.}

\subsection{Regret Testers}   \label{ss:regret-testers}

In the MARL algorithms surveyed above, the agents update all relevant parameters after each system interaction, and the multi-timescale analysis is implemented by controlling the magnitude of these every-step updates via step-size constraints. An alternative approach is to update certain iterates after each system interaction while updating other iterates infrequently, with many system interactions between updates. Such is the approach taken by works in the tradition of \emph{regret testing}, pioneered by Foster and Young in \cite{foster2006regret}, which presented an algorithm for stateless games. This approach was later studied by \cite{germano2007global} and \cite{Marden2009payoff} among others in the context of stateless repeated games, where impressive convergence guarantees were proven. The strong results of \cite{germano2007global} and \cite{Marden2009payoff} in stateless games were enabled to a significant extent by the absence of state dynamics, which would otherwise complicate value estimation and policy evaluation in multi-state games. 

In \cite{arslan2017decentralized},  the regret testing paradigm of Foster and Young was modified for multi-state stochastic games, where one must account for both the immediate cost of an action and also the cost-to-go, which depends on the evolution of the state variable. The \textit{decentralized Q-learning} algorithm of \cite{arslan2017decentralized} instructs agents to agree on an increasing sequence of policy update times, $( t_k )_{k \geq 0}$, and to fix one's policy within the intervals $[t_k, t_{k+1} )$, called the $k^{th}$ \emph{exploration phase}.\footnote{Although the terminology of exploration phases first appeared in \cite{arslan2017decentralized}, the exploration phase approach was pioneered by \cite{foster2006regret}.} In so doing, the joint policy process is fixed over each exploration phase, and within each exploration phase, each agent faces a Markov decision problem (MDP). Each agent then runs a standard Q-learning algorithm within the $k^{th}$ exploration phase, and updates its policy in light of its Q-factors at the end of the $k^{th}$ exploration phase. Since each agent faces an MDP,  one can analyze the learning iterates using single-agent learning theory. Provided the exploration phase lengths $T_k := t_{k+1} - t_k$ are sufficiently long, the probability of correctly recovering one's Q-function (defined with respect to the policies of the other agents and elaborated in the coming sections) can be made arbitrarily close to 1. 

In many of the papers on regret testers, the proofs of convergence to equilibrium share the same core elements, which can be summarized as follows. First, one defines an equilibrium event. (In the fully synchronized regime described above, this equilibrium event is usually the event that the joint policy of the $k^{th}$ exploration phase is an equilibrium.) Second, one uniformly lower bounds the probability of transiting from non-equilibrium to equilibrium in a fixed number of steps. Third, one shows that the probability of remaining at equilibrium once at equilibrium can be made arbitrarily large relative to the aforementioned lower bound. Fourth and finally, one obtains a high probability guarantee of equilibrium by analyzing the ratio of entry and exit probabilities to equilibrium policies (see, for instance, \cite[Lemma 2]{foster2006regret}).

In effect, the exploration phase technique decouples learning from adaptation, and allows for separate analysis of learning iterates and policy dynamics. This allows for approximation arguments to be used, in which the dynamics of the policy process resemble those of an idealized process where players obtain noise-free learning iterates for use in their policy updates. This has lead to a series of theoretical contributions in MARL that make use of the exploration phase technique, including \cite{yongacoglu2022decentralized} and \cite{yongacoglu2023satisficing}.
 
One natural criticism of the exploration phase technique described above is its reliance on synchronization of policy updating. In the description above, agents agree on the policy update times $\{ t_k \}_{k \geq 0}$ \textit{exactly}, and no agent ever updates its policy in the interval $(t_k, t_{k+1} - 1]$. Indeed, the mathematically convenient assumption of synchronization is made in various works in the regret testing tradition, including \cite{foster2006regret, germano2007global} and \cite{Marden2009payoff}. This can be justified in some settings, but is demanding decentralized settings where parameters are \blue{selected locally by each player or in distributed settings where parameter choices are communicated over noisy channels and are likely to differ across agents.}

To relax the synchronization assumption of regret testers, one allows each player $i$ to select its own exploration phase times $( t^i_k )_{k \geq 0}$. Player $i$ then has the opportunity to revise its policies at each time in $( t^i_k )_{ k \geq 0}$. We observe that an update time for player $i$ may fall within an exploration phase for some other player $j$. That is, $ t^j_m < t^i_k < t^j_{m+1}$ for two players $ i, j $ and exploration phase indices $k$ and $m$. In this case, the environment faced by player $j$ during its $m^{th}$ exploration phase is non-stationary, having been disrupted (at least) by player $i$'s changing policies. 

Intuitively, unsynchronized policy updating may be problematic for regret testers because the action-value estimates of a given player depend on historical data from that player's most recent exploration phase, which in turn depends on the (time-varying) joint policies used during this period. As such, if other players change their policies during an individual's exploration phase, the individual receives feedback from different sources, and its learning iterates may not approximate any quantity relevant to the prevailing environment at the time of the individual's policy update.  These changes of policy during an exploration phase constitute potential disruptions of a player's learning, and analysis of the overall joint policy process is difficult when players do not reliably learn accurate action-values. 

In this unsynchronized regime, learning and adaptation are not decoupled, as agents do not face stationary MDPs during each of their exploration phases. In particular, it is not guaranteed that a given player's Q-factors will converge. Consequently, it is not clear that the second and (especially) the third steps of the standard line of proof described above can be reproduced in the unsynchronized setting.

In \cite{Marden2009payoff}, a heuristic argument suggested that the use of inertia in policy updating may allow one to relax the assumption of perfect synchronization in regret testing algorithms for stateless repeated games. This argument offers a promising starting point, but also has two potential gaps. The first unresolved aspect pertains to the ratio of entry and exit probabilities. The heuristic argument correctly states that one can use inertia to both (i) lower bound the probability of transiting to equilibrium from out of equilibrium, and (ii) make the probability of remaining at equilibrium arbitrarily close to 1. However, the analysis of the overall joint policy process depends on the ratio of these entry and exit probabilities, which depends on the choice of inertia parameter and may be somewhat challenging to analyze. 

A second unresolved aspect has to do with the learning routine and how quickly errors are corrected. Here, the specific learning rate used in each agent's Q-learning algorithm plays an important role. Existing regret testing algorithms employ a decreasing learning rate, so as to ensure the convergence of learning iterates within a synchronized exploration phase. In the unsynchronized regime, however, agents learn against time-varying environments. If an agent uses a decreasing learning rate, it gives relatively high weight to outdated data that are no longer representative of its current environment. As such, it may take a long, uninterrupted stretch of learning against the prevailing environment to correct the outdated learning. The time required will depend on how much the learning rate parameters have been reduced, and these considerations must be quantified explicitly in the analysis. This point is somewhat subtle, and may pose a challenge if one attempts to use decreasing learning rates in the unsynchronized regime.

In this paper, we formalize the heuristic argument of \cite{Marden2009payoff} and show that it is effectively correct, with minor caveats. To rectify the unaddressed aspects mentioned above, we modify the Q-learning protocol to use a constant learning rate and conduct a somewhat involved analysis of the resulting behaviour. We focus on weakly acyclic games and best-response-based policy updating routines, but the analysis from this setting can be used to study other policy update routines and classes of games, including $\epsilon$-satisficing policy update dynamics such as those described in \cite{yongacoglu2023satisficing}.

\section{Unsynchronized Decentralized Q-Learning} \label{sec:learning-algorithms}

In this section, we present Algorithm~\ref{algo:main}, an unsynchronized variant of the decentralized Q-learning of \cite{arslan2017decentralized}. Unlike in the original decentralized Q-learning algorithm, Algorithm~\ref{algo:main} allows for the sequence of exploration phase lengths $\{T^i_k \}_{k \geq 0}$ vary by agent. \blue{Although this change introduces non-stationarity and thus complicates the analysis, the algorithm itself is no more intricate than that of \cite{arslan2017decentralized}.} 

\begin{algorithm2e}[h] 

	\SetAlgoLined
	\DontPrintSemicolon
	\SetKw{Receive}{Receive}
	\SetKw{Reset}{Reset}
	\SetKw{parameters}{Set Parameters}
	\SetKw{initialize}{Initialize}
 
	\SetKwBlock{For}{for}{end}
	
	\parameters \; 
	\Indp 
	$\{ T^i_k \}_{k \geq 0}$: player $i$'s sequence in $\nn$ of learning phase lengths \; 
	\vspace{1pt}
		\Indp
		Put $t^i_0 = 0$ and $t^i_{k+1} = t^i_k + T^i_k$ for all $k \geq 0$.  \; 
		\Indm
	\vspace{1pt}
	$\rho^i \in (0,1)$: experimentation probability \; 
	$\lambda^i \in (0,1)$: inertia during policy update \; 
	$\delta^i\in(0,\infty)$: tolerance level for suboptimality \; 
	$\alpha^i \in (0,1)$: step-size parameter (also called the learning rate) \; 
	\Indm
	\BlankLine
	
 	\initialize  $\pi_0^i \in \IndividualPolicySet^i_{SD}$, $\widehat{Q}_0^i \in \rr^{ \xx \times \IndividualActionSet^i }  $ (arbitrary) \\

	\For($k \geq 0$ ($k^{th}$ exploration phase for agent $i$ {)}){ 
		\For( $t = t_k^i, t_k^i +1, \dots, t^i_{k+1} - 1$)  
		{
			Observe $x_t$ \; 
			Select $a^i_t = \begin{cases} \pi^i_k ( x_t ) , &\text{w.p. } 1 - \rho^i  \\ \tilde{u}^i_t \sim \uniform(\IndividualActionSet^i), &\text{w.p. } \rho^i \end{cases} 	$  \;
			
			Observe cost $c^i_t := c( x_t, \ba_t )$, state $x_{t+1}$ \;
			
			Put $\Delta^i_t = c^i_t + \discount^i \min_{a^i} \widehat{Q}^i_t ( x_{t+1}, a^i )$ \;

			\vspace{3pt}
			$\widehat{Q}^i_{t+1} (x_t  ,a^i_t)  =  (1-\alpha^i ) \widehat{Q}_t^i(   x_t  , a_t^i) + \alpha^i  \Delta^i_t$  \;
			
			$\widehat{Q}_{t+1}^i(x,u^i) =  \widehat{Q}^i_t (x,u^i)$,   for all $(x,u^i)\not=(x_t, a_t^i)$\;

		}
		\vspace{3pt}  
		
		\BlankLine 
	
		\If { $\pi^i_k \in \widehat{\BR}^i_{ \delta^i} ( \widehat{Q}^i_{t^i_{k+1}} )$,}{ $\pi^i_{k+1} \leftarrow \pi^i_k$ }
		\Else( ){$\pi^i_{k+1} \leftarrow \begin{cases} \pi^i_k , &\text{w.p. } \lambda^i \\ \tilde{\pi}^i_k \sim \uniform \left(    \widehat{\BR}^i_{  \delta^i} ( \widehat{Q}^i_{t^i_{k+1}} )  \right) , &\text{w.p. } 1 - \lambda^i \end{cases} $ }
 
	}
	
	\caption{Unsynchronized Decentralized Q-Learning} \label{algo:main}
\end{algorithm2e}

\subsection{Assumptions}

In order to state our main result, \Cref{theorem:main}, we now impose some assumptions on the underlying game $\GG$, on the algorithmic parameters, and on the mutual independence of certain randomized subroutines appearing in the algorithm.

\subsubsection{\blue{Assumption on the transition kernel}}

\begin{assumption} \label{assumption:state-visitation} 
For any pair $(s, s') \in \xx \times \xx$, there exists $H = H( s, s' ) \in \nn$ and a sequence of joint actions $\ba^{\prime}_0, \dots, \ba^{\prime}_H \in \JointActionSet$ such that 
\[
\P ( x_{H+1} = s' | x_0 = s, \ba_0 = \ba^{\prime}_0, \dots, \ba_H = \ba^{\prime}_H ) > 0 .
\]
\end{assumption}

\vspace{5pt}
\Cref{assumption:state-visitation} requires that the state process can be driven from any initial state to any other state in finitely many steps, provided a suitable selection of joint actions is made. This is a rather weak assumption on the underlying transition kernel $P$, and is quite standard in the theory of MARL (c.f. \cite[Assumption 4.1, Case iv]{ozdaglar2021independent}).

\subsubsection{\blue{Assumptions on algorithmic parameters}}

Our next assumption restricts the hyperparameter selections in Algorithm~\ref{algo:main}. Let $\bar{\delta} := \min \left( \mathfrak{A} \setminus \{ 0 \} \right)$, where
\begin{align*}
 \mathfrak{A}  :=  {\big \{}  &  {\big \vert} Q^{*i}_{ \bpi^{-i}} ( s, a^i_1 ) - Q^{*i}_{\bpi^{-i}} ( s, a^i_2)  {\big \vert } :   i \in \NN, \bpi^{-i} \in \JointPolicySet^{-i}_{SD}, s \in \xx,  a^i_1, a^i_2 \in \IndividualActionSet^i {\big \} } .
 \end{align*}

The quantity $\bar{\delta}$, defined originally by \cite{arslan2017decentralized} and recalled above, is the minimum \textit{non-zero} separation between two optimal Q-factors with matching states, minimized over all agents $i \in \NN$ and over all policies $\bpi^{-i} \in \JointPolicySet^{-i}_{SD}$.

For any baseline policy $\bpi \in \JointPolicySet_{SD}$ and fixed experimentation parameters $\{ \rho^i \}_{i \in \NN}$, we use the notation $\hat{\bpi} \in \JointPolicySet_{S}$ to denote a corresponding behaviour policy, which is stationary but not deterministic. When using $\hat{\pi}^i$, agent $i \in \NN$ follows $\pi^i$ with probability $1- \rho^i$ and mixes uniformly over $\IndividualActionSet^i$ with probability $\rho^i$. It was previously shown in \cite[Lemma B3]{arslan2017decentralized} that the optimal Q-factors for these two environments will be close provided $\rho^i$ is sufficiently small for all $i \in \NN$. In particular, there exists $\bar{\rho} > 0$ such that if $\rho^i \in (0, \bar{\rho})$ $\forall i \in \NN$, then 
\begin{equation}
\| Q^{*j}_{\bpi^{-j}} - Q^{*j}_{\hat{\bpi}^{-j}} \|_{\infty} < \frac{ \min_{i \in \NN} \min \{ \delta^i, \bar{\delta} - \delta^i \}}{4}  , \: \forall j \in \NN, \bpi^{-j} \in \JointPolicySet^{-j}_{SD} .   \label{eq:hat-policy}
\end{equation}

\vspace{2.5pt} 

\begin{assumption} \label{assumption:rho-delta}
For all $i \in \NN$, $\delta^i \in ( 0, \bar{\delta} )$ and $\rho^i \in ( 0, \bar{\rho} )$. 
\end{assumption}

\vspace{2.5pt} 

In the first part of \Cref{assumption:rho-delta}, we require that $\delta^i$, player $i$'s tolerance for suboptimality in its Q-factor, is positive to account for noise in its learning iterates, as exact convergence is unreasonably demanding during a truncated learning phase. On the other hand, the tolerance parameter $\delta^i$ is upper bounded so as to exclude truly suboptimal policies from consideration in an estimated best-response set. The second part of \Cref{assumption:rho-delta} concerns the experimentation parameter $\rho^i$. We assume that $\rho^i > 0$ so that player $i$ can explore all of its actions and accurately learn its Q-factors in the limit. This experimentation parameter $\rho^i$ must be kept small, however, so player $i$'s behaviour policy closely approximates its baseline policy.  \Cref{assumption:rho-delta} appears as Assumption~2 in \cite{arslan2017decentralized}.

\vspace{2.5pt} 

%
\begin{assumption} \label{assumption:exploration-phases}
There exist integers $R, T \in \nn$ such that $T \leq T^i_k \leq RT$ for each player $i$ and exploration phase index $k \geq 0$. 
\end{assumption}

\Cref{assumption:exploration-phases} requires that each agent spends a long time learning during any given exploration phase ($T^i_k \geq T$ almost surely for any player $i$ during any exploration phase $k$). The second part of \Cref{assumption:exploration-phases}, that $T^i_k \leq RT$ almost surely for all players $i$ and exploration phase indices $k$, is assumed in order to control the relative frequency at which different agents update their policies. This is done to avoid the potentially problematic regime in which some agents update their policies an enormous number of times and repeatedly disrupt the learning of another agent who uses long exploration phase lengths.

\vspace{5pt}

When all players use Algorithm~\ref{algo:main}, we view the resulting sequence of policies $\{ \pi^i_k \}_{k \geq 0}$ as player $i$'s \emph{baseline} policy process, where $\pi^i_k$ is player $i$'s baseline policy during $[t^i_k, t^i_{k+1})$, player $i$'s $k^{th}$ exploration phase. This policy process $\{ \pi^i_k \}_{k \geq 0}$ is indexed on the coarser timescale of exploration phases. For ease of reference, we also introduce a sequence of baseline policies indexed by the finer timescale of stage games. For $t \geq 0$ with $t \in [t^i_k, t^i_{k+1} )$, we let $\phi^i_t := \pi^i_k$ denote player $i$'s baseline policy during the stage game at time $t$. The baseline joint policy at stage game $t$ is then denoted $\bphi_t = ( \phi^i_t )_{i \in \NN}$. Furthermore, we refer to the collection of Q-factor step-size parameters $\{ \alpha^i \}_{i \in \NN} $ as $\balpha \in (0,1)^N$.

\subsubsection{Assumptions on algorithmic randomness} \label{ss:primitive-random-variables}
\blue{Algorithm~\ref{algo:main} is a randomized algorithm: in multiple places (such as Lines 12 and 20), a player's decision depends on both the history of play and also on exogenous randomness. In this section, we describe our assumptions on how these random selections are made. In simple terms, we assume that random choices are independent across players and are conditionally independent of the history of the system. The intuition is elaborated below and then formalized in \Cref{assumption:primitive-random-variables} using the language of primitive random variables.} 

\vspace{5pt}

\blue{Consider the action choice in Line 12 of Algorithm~\ref{algo:main}:}

\blue{
\[
\text{Select } a^i_t = \begin{cases} 	\pi^i_k ( x_t ) , 			&\text{w.p. } 1 - \rho^i  \\ 
							\tilde{u}^i_t \sim \uniform(\IndividualActionSet^i), 	&\text{w.p. } \rho^i 
				\end{cases} 	
\]
}

\blue{The selection of action $a^i_t$ can be expressed as a function of the state $x_t$ along with the outcome of two random quantities: a uniform random variable, denoted $\tilde{\rho}^i_t \sim \uniform( [0,1])$, and a uniform random action choice $\tilde{u}^i_t \sim \uniform ( \IndividualActionSet^i )$. That is, Line 12 of Algorithm~\ref{algo:main} can be equivalently expressed as
\begin{equation}
a^i_t   =	\begin{cases}
			\pi^i_k ( x_t ) , 	&\text{if } \tilde{\rho}^i_t \geq 1- \rho^i  \\
			\tilde{u}^i_t ,	&\text{if } \tilde{\rho}^i_t < \rho^i . 
		\end{cases}  \label{eq:primitive-rho-action}
\end{equation}
}

\blue{
Similarly, the selection of the policy $\pi^i_{k+1}$ in Line 20 of Algorithm~\ref{algo:main} can be expressed as a function of player $i$'s Q-factor estimates and the outcome of two random quantities: }
\begin{itemize}
	\color{black}
	
	\item A uniform random variable, denoted $\tilde{\lambda}^i_t \sim \uniform( [0,1])$.
	
	\item A uniform random policy $\tilde{\pi}^i_k := \tilde{\pi}^i_k ( B^i ) \sim \uniform ( B^i )$, where $B^i = \widehat{\BR}^i_{\delta^i} \left( \widehat{Q}^i_{t^i_{k+1}}	\right)$.
\end{itemize}

\blue{
\noindent That is, Line 20 of Algorithm~\ref{algo:main} can be equivalently expressed as
\[
\pi^i_{k+1} = 	\begin{cases}
				\pi^i_k, &\text{if } \tilde{\lambda}^i_k < \lambda^i	\\
				\tilde{\pi}^i_k  , &\text{else.}
			\end{cases}
\]
}
\blue{
In the analysis of the coming sections, it will be convenient to isolate the role of exogenous randomness and separate it from the realization of the Q-factor iterates above. It will therefore be convenient to consider analogs of the random choice $\tilde{\pi}^i_k (B^i) $ drawn from arbitrary policy subsets $B^i \subseteq \IndividualPolicySet^i_{SD}$.
}

\vspace{11pt} 

With the preceding intuition in mind, we now introduce several collections of \emph{primitive} random variables that will be used to state \Cref{assumption:primitive-random-variables} below. For any player $i \in \NN$ and $t \geq 0$, we define the following random variables:

\begin{itemize}
	
	\setlength \itemsep{2.5pt}
	
	\item $\{ W_t \}_{t \geq 0}$ is an identically distributed, $[0,1]$-valued stochastic process. For some $f : \xx \times \JointActionSet \times [0,1] \to \xx$, state transitions are driven by 
	$\{ W_t   \}_{t \geq 0}$ via $f$: 
		\begin{align*}
			\P ( x_{t+1} = s' | x_t = s, \ba_t = \ba ) = \P \left( W_t  \in \{ w : f( s, \ba, w ) = s' \} \right), 
		\end{align*}
	
	for any $(s,\ba, s') \in \xx \times \JointActionSet \times \xx$ and $ t \geq 0$;
		
	\item $\tilde{u}^i_t \sim \uniform( \IndividualActionSet^i )$;
	
	\item $\tilde{\rho}^i_t \sim \uniform( [0,1])$;
	
	\item $\tilde{\lambda}^i_t \sim \uniform( [0,1] )$;
	
	\item For non-empty $B^i \subseteq \IndividualPolicySet^i_{SD}$, $\tilde{\pi}^i_t ( B^i ) \sim \uniform ( B^i )$;
	
\end{itemize}

\begin{assumption} \label{assumption:primitive-random-variables}
The collection of primitive random variables $\VV_1 \cup \VV_2$ is mutually independent, where 
\begin{align*}
\VV_1 &:= \bigcup_{i \in \NN, t \geq 0 } \left\{ W_t, \tilde{\rho}^i_t, \tilde{u}^i_t, \tilde{\lambda}^i_t    \right\},    				\\
\VV_2 &:= \bigcup_{i \in \NN, t \geq 0} \left\{ \tilde{\pi}^i_t ( B^i ) : B^i \subseteq \IndividualPolicySet^i_{SD}, B^i \not= \varnothing \right\}. 
\end{align*}
\end{assumption}

\noindent \textbf{Remark:} The random variables in $\VV_1 \cup \VV_2$ are primitive random variable that do not depend on any player's choice of hyperparameters or on the history of play of the game. 

\subsection{Theorem Statement} \label{ss:theorem-statement}

\begin{theorem} \label{theorem:main}
Let $\GG$ be a weakly acyclic game and suppose each player $i \in \NN$ uses Algorithm~\ref{algo:main} to play $\GG$. Suppose Assumptions~\ref{assumption:rho-delta}, \ref{assumption:exploration-phases}, and  \ref{assumption:primitive-random-variables}  hold, and let $\epsilon > 0$. There exists $\bar{\alpha}_{\epsilon} > 0 $ and a function $\bar{T}_{\epsilon} : (0,1)^N \times \nn \to \nn$ such that if 
\[
\max_{i \in \NN} \alpha^i < \bar{\alpha}_{\epsilon} , \text{ and } T \geq \bar{T}_{\epsilon} ( \balpha, R ) ,
\]
then $ \P ( \bphi_t \in \eq[0]_{SD} ) \geq 1 - \epsilon,$ for all sufficiently large $t \in \nn$. 
\end{theorem}


\ 

In words, \Cref{theorem:main} says the following: if the relative frequency of policy update times of different players is controlled by some ratio $R$, if players use sufficiently small learning rates $\balpha$, and if the lower bound $T$ on each player's exploration phase length is sufficiently long relative to both the learning rates $\balpha$ as well as the ratio parameter $R$, then play will be driven to equilibrium with high probability.

\vspace{5pt}

As we will see, the interdependence of the various parameters---$R, \balpha,$ and $T$---is significant in the analysis to follow. 

\section{Proof Overview}   \label{sec:proof-outline}
\blue{
As we will see in \Cref{sec:proof}, the proof of \Cref{theorem:main} is somewhat intricate. For this reason, we now offer an outline of the proof to facilitate its reading. 
}

\subsection{High-Level Overview}

\blue{
The proof of \Cref{theorem:main} involves three major steps. First, we introduce a sequence of equilibrium events, $\{ B_k \}_{k \geq 0}$, defined in terms of a suitably defined sequence of time intervals $\{ [ \tau^{\min}_k, \tau^{\max}_k ] \}_{k \geq 0}$, to be elaborated in the sequel. For $k \geq 0$, we put
\[
B_k := \left\{ \bphi_t =  \bphi_{\tau^{\min}_k} \in \eq_{SD}: t = \tau^{\min}_k + 1, \cdots, \tau^{\max}_k \right\} .
\]

In words, $B_k$ is the event in which the baseline policy did not change during the interval $[\tau^{\min}_k , \tau^{\max}_k]$ and moreover the baseline policy was an equilibrium during this time.
}

\blue{
Second, for a particular integer $L < \infty$ (defined in the text preceding \Cref{lemma:stay-at-equilibrium}), we argue in \Cref{lemma:stay-at-equilibrium} that we can control and lower bound the probability of remaining at the equilibrium event for $L$ steps: that is, $\P ( B_{k+L} | B_k ) $ can be made arbitrarily large by suitable selection of algorithm parameters. The proof of \Cref{lemma:stay-at-equilibrium} uses convergence properties of Q-learning iterates (\Cref{lemma:primitive-random-variables-lead-to-Q-convergence}) and requires careful arguments about conditional probabilities, since conditioning on the event $B_k$ carries confounding information about the state-action trajectory from both \emph{before and after} time $\tau^{\min}_k$ and precludes simple analysis that disconnects the future evolution of Q-learning estimates from the history of the system before time $\tau^{\min}_k$. Consequently, the analysis is somewhat delicate, and is worded in terms of primitive random variables, which offers a path around the confounding effect of conditioning on $B_k$.}

\blue{
Third, we argue in \Cref{lemma:go-to-equilibrium} that the probability of driving play to this equilibrium event in $L$ time steps can be lower bounded by a positive constant.  After this is achieved, one can then explicitly lower bound $\P ( B_{k+mL} )$ for suitably large $m$, as in \cite{arslan2017decentralized} and \cite{yongacoglu2023satisficing}.
}

\section{Proof of \cref{theorem:main}} \label{sec:proof}


\subsection{Additional Notation and Constructions}   \label{ss:more-notation} We begin by defining several objects that will be needed for the various lemmas to follow. \blue{The notation in this section builds on that of \Cref{ss:primitive-random-variables}. For ease of reference, we have included a glossary of notation in \Cref{table:algo-notation}, \Cref{table:analysis-notation}, and \Cref{table:analysis-notation-continued} of \Cref{appendix:notation}.}

\vspace{5pt}

\blue{
Recall that in  \Cref{ss:primitive-random-variables}, we motivated the use of primitive random variables for describing the algorithmic randomness of Algorithm~\ref{algo:main}. The primitive random variables described earlier included $\{ \tilde{u}^i_t , \tilde{\rho}^i_t \}_{t \geq 0}$, which appeared in randomized action selection, and $\{ \tilde{\lambda}^i_k , \tilde{\pi}^i_k \}_{k \geq 0}$, which appeared in randomized policy updated. We also recall that $\tilde{\pi}^i_k$ was previously a random sample from a particular policy subset $B^i \subseteq \IndividualPolicySet^i_{SD}$, which was defined using player $i$'s Q-factor iterates. In this section, it will be helpful to further isolate the role of algorithmic randomness, and therefore we will also consider random samples from all non-empty subsets of $\IndividualPolicySet^i_{SD}$. For the latter purpose and} for each $i \in \NN$, we arbitrarily  order non-empty subsets of $\IndividualPolicySet^i_{SD} $ as $B^{i,1} ,\dots, B^{i, m_i}$, where $m_i = | 2^{\IndividualPolicySet^i_{SD}} \setminus \{ \varnothing \} |$. 

\vspace{5pt} 
\noindent We introduce the following new quantities for each $t \geq 0$:

\vspace{2.5pt}

\begin{itemize}
	\setlength \itemsep{2.5pt} 
	
	\item $ \omega^i_t := ( \tilde{\rho}^i_t, \tilde{u}^i_t, \tilde{\lambda}^i_t, \tilde{\pi}^i_t ( B^{i,1}), \dots, \tilde{\pi}^i_t ( B^{i, m_i} ) ),$ for all $i \in \NN$. We define $\mathfrak{S}^i$ to be the set in which the random quantity $\omega^i_t$ takes its values. That is, 
		\[
			\mathfrak{S}^i := [0,1] \times \IndividualActionSet^i \times [0,1] \times B^{i,1} \times \cdots \times B^{i,m_i } , \quad \forall i \in \NN. 
		\]  
		
	\item[] \blue{In plain terms, $\omega^i_t$ consists of random samples of each of the quantities needed for player $i$ to execute the randomized subroutines in Algorithm~\ref{algo:main} at a given time $t$.}
	
	\item $\bomega_t := ( W_t, \omega^1_t, \dots, \omega^N_t )$, where $\{W_t \}_{t \geq 0}$ is the $[0,1]$-valued independent and identically distributed (i.i.d.) noise process driving state transitions that was introduced in \S\ref{ss:primitive-random-variables}.
	
	\item[] \blue{For each $t$, $\bomega_t$ is a collection of the randomized quantities described in $\omega^i_t$ along with the random noise that drives state transitions. Of note, the information contained in $\bomega_t$ is sufficient for recovering  the subsequent state $x_{t+1}$, the current joint action $\ba_t$, and the subsequent joint policy $\bphi_{t+1}$.}
	
	\item $\bvarpi_{t\blue{+}} := ( \bomega_t, \bomega_{t+1}, \dots )$. \blue{This quantity describes all random samples required to fully describe the future of the state-action trajectory and thus of Q-factors.}
	
	\item $\widehat{\bQ}_t := ( \widehat{Q}^1_t, \dots, \widehat{Q}^N_t )$. 
	
	\item \blue{$\bh_t$} $:= ( x_0 , \bphi_0 , \widehat{\bQ}_0, \dots,  x_t , \bphi_t , \widehat{\bQ}_t )$. \blue{This quantity describes the states, joint policies, and Q-factors up to and including time $t$.}
	
	\item \blue{$\bH_t$} $:= \left( \xx \times \JointPolicySet_{SD} \times \rr^{\xx \times \IndividualActionSet^1} \times \dots \times \rr^{\xx \times \IndividualActionSet^N} \right)^{t+1}$. 
	
	\item \blue{$\bH_{t, \rm eq} $} $:= \{ \bh_t \in \bH_t : \bphi_t \in \eq[0]_{SD} \}$. \blue{This set consists of those $t$-histories for which the final joint policy $\bphi_t$ is an equilibrium.}
	
\end{itemize}

\

We note that $\omega^i_t$ is a random quantity taking values in $\mathfrak{S}^i$, while $\bomega_t$ is a random quantity taking values in $[0,1] \times \mathfrak{S}^1 \times \cdots \times \mathfrak{S}^N$. For each $t \geq 0$, a realization of the random quantity $\bvarpi_{t+}$ is a sequence with entries in $[0,1] \times \mathfrak{S}^1 \times \cdots \times \mathfrak{S}^N$.

\

When every player uses Algorithm~\ref{algo:main} to play the game and each player's randomization mechanism is governed by the primitive random variables of \S\ref{ss:primitive-random-variables}, we have that the sample path of play---including realizations of the state process $\{x_t \}_{t \geq 0}$, the action processes $\{ a^i_t \}_{t \geq 0}$ for each agent $i \in \NN$, the baseline joint policy process $\{ \bphi_t \}_{t \geq 0}$, and the Q-factor estimates $\{ \widehat{Q}^i_t \}_{t \geq 0}$---are deterministic functions of any given prefix history $\bh_s \in \bH_s$ and its corresponding continuation, $\bvarpi_{s+} \in \left( [0,1] \times \mathfrak{S}^1 \times \cdots \times \mathfrak{S}^N \right)^{\infty}$, where $s \in \zz_{\geq 0}$. With this in mind, for $i \in \NN$ and $t \geq 0$, we define mappings $\QQ^i_t $ and $\Phi_t $ such that 
\begin{equation}
\widehat{Q}^i_t = \QQ^i_{t} ( \bh_s, \bvarpi_{s+}  ), \: \text{ and } \: \bphi_t = \Phi_t ( \bh_s, \bvarpi_{s+} ), \quad \forall 0 \leq s \leq t . \label{def:Q-and-Phi-mappings}
\end{equation} 

%

Next, for any $s,t \geq 0$ and any $i \in \NN$, we introduce a function $\bar{Q}^i_{t+s} ( \bh_t, \bvarpi_{t+} )$ that reports the \emph{hypothetical Q-factors} player $i$ \emph{would have} obtained if the baseline policies had been frozen at time $t$. That is, the history up to time $t$ is given by $\bh_t$, the primitive random variables from $t$ onward are given by $\bvarpi_{t+}$, and we obtain the hypothetical ($\bvarpi_{t+}$-measurable) continuation trajectory $( \bar{x}_t, \bar{\ba}_t, \dots, \bar{x}_{t+s}, \bar{\ba}_{t+s})$, as $\bar{x}_t := x_t $,
\[
\bar{x}_{t+m+1} = f ( \bar{x}_{t+m}, \bar{\ba}_{t+m}, W_{t+m} ), \quad \forall \, 0 \leq m \leq s,
\]
where for each player $j$ and time $t+m \geq t$, 
\[
\bar{a}^j_{t+m} :=	\begin{cases}
				\phi^j_{t} ( \bar{x}_{t+m} ) , &\text{if } \tilde{\rho}^j_{t+m} > \rho^j \\
				\tilde{u}^j_{t+m} &\text{otherwise.}
				\end{cases}
\]

Note that the index of $\phi^i_{t}$ is not $t+m$, which reflects that, in this hypothetical continuation, the baseline policies were frozen at time $t$. Each hypothetical Q-factor estimate $\bar{Q}^i_{t+s} ( s,a^i )$ is then (analytically) built out of this hypothetical trajectory using the regular Q-learning update with the initial condition prescribed by $\widehat{\bQ}_t$.

\

\noindent\textbf{Remark:} Hypothetical Q-factors feature prominently in the analysis of the coming sections, and are a somewhat subtle construction, so we explain their utilization here. To analyze the likelihood of driving play to equilibrium or the likelihood of remaining at equilibrium, we ultimately wish to make probabilistic statements about the realized Q-learning iterates, $\{ \widehat{\bQ}_t \}_{t \geq 0}$, conditional on various events, such as the event that players did not switch their baseline policies recently. However, conditioning on such events can be problematic in the analysis, since the conditioning event clearly carries information about the state-action trajectory that may be consequential for the evolution of the Q-factor estimates. Due to this confounding effect, we will avoid making statements about conditional probabilities directly, and we will instead study trajectories of the hypothetical Q-factors. In so doing, we can describe likelihoods of relevant events in terms of the primitive random variables, thereby avoiding the confounding effects of conditioning.

\subsection{Supporting Results on Q-Learning Iterates}   \label{ss:supporting-results}

\begin{lemma} \label{lemma:uniform-boundedness}
For some $M < \infty$, the following holds almost surely:
\[
\max_{i \in \NN} \sup_{t \geq 0} \left\| \widehat{Q}^i_t \right\|_{\infty}  \leq M. 
\]
\end{lemma}

\vspace{5pt}

\noindent 

\begin{proof}
For all $ i \in \NN$, we have 
\begin{align*}
\| \widehat{Q}_{t+1}^i \|_{\infty} & \leq \max \bigg\{ (1-\alpha^i )\| \widehat{Q}_{t}^i\|_{\infty} +  \alpha^i \big(\|c^i\|_{\infty} + \discount^i \| \widehat{Q}_{t}^i\|_{\infty}\big),\| \widehat{Q}_{t}^i\|_{\infty} \bigg\}.
\end{align*}
Defining $M := \max_{i \in \NN} \left\{ \left\| \widehat{Q}^i_0 \right\|_{\infty} , \frac{  \| c^i \|_{\infty}  }{1 - \discount^i } \right\}$, we have that if $\left\| \widehat{Q}^i_t \right\|_{\infty} \leq M$, then
\begin{align*}
\| \widehat{Q}_{t+1}^i\|_{\infty} & \leq \max\bigg\{ (1-\alpha^i ) M +  \alpha^i \big(\|c^i\|_{\infty} + \discount^i M \big), M \bigg\} \\
& = \max\bigg\{ M +  \alpha^i \big(\underbrace{\|c^i\|_{\infty} - (1-\discount^i) M}_{\leq0} \big), M \bigg\}  =M .
\end{align*}
This proves the lemma, as $\| \widehat{Q}_0^i\|_{\infty} \leq M < \infty$ for each $i \in \NN$.
\end{proof}

\

The following lemma says that players can learn their optimal Q-factors accurately when they use sufficiently small step sizes and when their learning is not disrupted by policy updates for a sufficiently long number of steps. It is worded in terms of primitive random variables and hypothetical continuation Q-factors for reasons described above.

\vspace{5pt}

\begin{lemma}  \label{lemma:primitive-random-variables-lead-to-Q-convergence} 
For any $\xi>0$, there exists $\hat{\alpha}_{\xi}>0$ and function $\hat{T}_{\xi}:(0,1)^N \to \nn$ such that if (1) $\alpha^i \in (0, \hat{\alpha}_{\xi}) $ for all $ i \in \NN$, and  (2) $\tilde{T} \geq \hat{T}_{\xi}( \balpha)$, then
\[
\P \left(\Omega_{t:t+\tilde{T}} \right) \geq 1-\xi , \quad \forall \, t \geq 0,  \bh_t \in \bH_t ,
\]
where $\Omega_{t:t+\tilde{T}} = \big\{ \bvarpi_{t+} :\max_{i\in\NN} \| \bar{Q}_{t+\tilde{T}}^i(  \bh_t  , \bvarpi_{t+} )- {Q}_{\bphi_{t}^{-i}}^{*i}\|_{\infty}<\xi \big\}.$
\end{lemma}

\vspace{5pt}

\begin{proof} Since each player $i$'s policy is $\rho^i$-soft, our \Cref{assumption:state-visitation} implies the persistent excitation assumption of \cite[Assumption 1]{beck2012error}. The result then follows from \Cref{lemma:uniform-boundedness} and \cite[Theorem 3.4]{beck2012error}, using Markov's inequality.  
\end{proof}

In the unsynchronized regime, where each agent faces a non-stationary, time-varying environment and uses Q-learning, \Cref{lemma:primitive-random-variables-lead-to-Q-convergence} is an extremely useful result that allows one to quantify the amount of time needed to correct outdated information contained in the learning iterates. The choice to employ constant learning rates was made in large part due to this consideration. 

In the sequel, we fix $\hat{\alpha}_{\xi}$ and $\hat{T}_{\xi} (\cdot)$ with the properties outlined in \Cref{lemma:primitive-random-variables-lead-to-Q-convergence}.

\subsection{A Supporting Result Controlling Update Frequencies} \label{ss:active-phases}

A core challenge of analyzing unsynchronized multi-agent learning is that policy updates for one player constitute potential destabilizations of learning for others. We now recall the sequence of time intervals $\{ [ \tau^{\min}_k , \tau^{\max}_k ] \}_{k \geq 0 }$ previously described in \Cref{sec:proof-outline}. These time intervals will be useful in quantifying and analyzing the effects of such disruptions. In particular, we wish to upper bound the number of potential disruptions to a given player's learning, which can be upper bounded by the number of times other players finish an exploration phase during the given individual's current exploration phase. Producing an upper bound of this sort will be important for quantifying the importance of inertia for stabilizing the environment in the heuristic argument of \cite{Marden2009payoff}. 


\begin{definition}   \label{def:active-phases}
Let $\tau^{\min}_0 = \tau^{\max}_0 := 0$. For $k \geq 0$, define
\begin{align*}
\tau^{\min}_{k+1} 	&:= \inf \{ t^i_n : t^i_n > \tau^{\max}_{k} , i \in \NN, n \geq 0 \} \\
\tau^{\max}_{k+1}	&:= \inf {\big \{ } t \geq \tau^{\min}_{k+1} :  \forall \, i \in \NN, \exists \, n \in \nn \text{ s.t. } t^i_{n} \in [\tau^{\min}_{k+1} , t ],   \\  	
					&\qquad \quad \qquad \quad \qquad \text{ and } \inf \{  t^i_{\bar{n}} > t : i \in \NN, \bar{n} \in \nn \} \geq t + T/N { \big \}},
\end{align*}
where $T$ is the constant appearing in \Cref{assumption:exploration-phases}.
\end{definition}

\vspace{5pt} 

The intervals $[ \tau^{\min}_k , \tau^{\max}_k ]$ represent active phases, during which players may change their policies. The $(k+1)^{th}$ active phase starts at $\tau^{\min}_{k+1}$, which is defined as the first time after $\tau^{\max}_k$ at which some agent has an opportunity to revise its policy. As a consequence of these definitions, no policy updates occur in $( \tau^{\max}_k, \tau^{\min}_{k+1} )$. 

The definition of $\tau^{\max}_{k+1}$ is slightly more involved: $\tau^{\max}_{k+1}$ is the minimal stage game time at which (a) each agent has an opportunity to switch policies in $[\tau^{\min}_{k+1}, \tau^{\max}_{k+1}]$ and (b) that no agent finishes its current exploration phase in the next $T/N$ stage games. The latter point is critical, as it guarantees that each player's learning is allowed to proceed uninterrupted for at least $T/N$ stages after the final (potential) disruption.


\vspace{5pt} 

\begin{lemma} \label{lemma:active-phases} 
The sequences $\{ \tau^{\min}_{k} \}_{k \geq 0}$ and $\{ \tau^{\max}_{k} \}_{k \geq 0}$ are well-defined (i.e. the infimum defining each term is achieved by a finite integer), and the following holds for any $k \geq 0$:
\begin{itemize}
	\setlength \itemsep{3.5pt} 
	
	\item[(a)] $\tau^{\min}_{k+1} \geq \tau^{\max}_{k} + T/N$.
	
	\item[(b)] For each $i \in \NN$, we have \emph{$\sum_{n \geq 0} \textbf{1} \{ t^i_n \in [ \tau^{\min}_k , \tau^{\max}_k ] \} \leq R + 1.$}
\end{itemize}
\end{lemma}

\vspace{5pt} 
%
%

In words, part (a) of \Cref{lemma:active-phases} guarantees that any successive active phases are separated by at least $T/N$ stage games, while part (b) guarantees that any individual player has at most $R+1$ opportunities to revise its policy during the $k^{th}$ active phase.

\vspace{2.5pt} 

\begin{proof}
For some $k \geq 0$, suppose that (a) and (b) hold for all $l \leq k$. That is, 
\begin{itemize}
	\item[(a)] $\tau^{\min}_{l + 1} \geq \tau^{\max}_{l} + T/N$, for all $0 \leq l \leq k$, and
	\item[(b)] for each $i \in \NN, l \leq k$,  we have $\sum_{n \geq 0} \textbf{1} \{ t^i_n \in [ \tau^{\min}_l , \tau^{\max}_l ] \} \leq R+1. $
\end{itemize}

\

\noindent Note that this holds for $k = 0$. We will show the following: (1) $\tau^{\max}_{k+1} < \infty$; (2)~$\tau^{\min}_{k + 2}~\geq~\tau^{\max}_{k+1}~ +~T/N$; (3) For all $i \in \NN$, we have $\sum_{n \geq 0} \textbf{1} \{ t^i_n \in [ \tau^{\min}_{k+1}, \tau^{\max}_{k+1} ] \} \leq R+1. $

\

\noindent For each $i \in \NN$, let $n_i \in \zz_{\geq 0}$ denote the index such that 
\[
t^i_{n_i - 1} < \tau^{\min}_{k+1} \leq t^i_{n_i}.
\]

By minimality of $\tau^{\min}_{k+1}$, we have that $t^i_{n_i - 1} \leq \tau^{\max}_k$. By \Cref{assumption:exploration-phases}, we have $T^i_{n_i - 1} \leq RT$ and thus 
\[
t^i_{n_i} = t^i_{n_i - 1} + T^i_{n_i - 1 } \leq \tau^{\max}_k + RT.
\]

This implies $t^i_{n_i} - \tau^{\min}_{k+1} \leq RT - T/N$, since $\tau^{\min}_{k+1} \geq \tau^{\max}_{k} + T/N$ by hypothesis. 

\vspace{2.5pt}

We put $\hat{t}_0 := \max_i t^i_{n_i}$, and let $j(0) \in \NN$ denote an agent with $t^{j(0)}_{n_{j(0)}} = \hat{t}_0$.  

\vspace{2.5pt}

For each $l \in \{ 0, 1, \dots, N-1\}$, we define $\hat{t}_{l+1}$ as 
\[
\hat{t}_{l+1} = \min \left\{ t^j_{\bar{k}} > \hat{t}_l : j \in \NN, \bar{k} \geq 0  \right\}. 
\]

Observe that $\hat{t}_0 \leq \tau^{\max}_{k+1}$. Moreover, if there exists some $ l \leq N-1 $ such that $\hat{t}_{l+1} \geq \hat{t}_l + T/N$, then $\tau^{\max}_{k+1} \leq \hat{t}_l$, and thus $\tau^{\max}_{k+1} < \infty$. We now argue that such $l$ exists.

\vspace{2.5pt}

Suppose that no such $l$ exists: 
\begin{align}
			&\max \{ \hat{t}_1 - \hat{t}_0 , \cdots, \hat{t}_{N} - \hat{t}_{N-1} \} < T/N    \\ 
\Rightarrow 	&\hat{t}_N - \hat{t}_0 = \sum_{l = 0}^{N-1} \left( \hat{t}_{l+1} - \hat{t}_l  \right) < N \cdot \frac{T}{N}  = T  \leq \min_{i,n} T^i_n . \nonumber
\end{align}

One concludes that the minima defining each $\{ \hat{t}_l : 1 \leq l \leq N \}$ are attained by $N$ distinct minimizing agents, one of whom is $j(0)$. But then, for some $l$, we have 
\[
\hat{t}_0 = t^{j(0)}_{n_{j(0)}} , \text{ and } \: t^{j(0)}_{n_{j(0)}}  + T^{j(0)}_{n_{j(0)}} = \hat{t}_l \leq \hat{t}_N,
\]
which implies $T^{j(0)}_{n_{j(0)}} < T$, contradicting \Cref{assumption:exploration-phases}. 

\vspace{2.5pt}


We have thus shown that the set $\mathfrak{T} \not=\varnothing$, where $\mathfrak{T}$ is given by
\[
\mathfrak{T} := \{ l \in \{0, 1, \dots, N-1\} : \hat{t}_{l+1} \geq \hat{t}_{l} + T/N \}. 
\]

Let $l^{*} = \min \mathfrak{T}$, and note that, if $l^{*} \not=0$, then $\hat{t}_{k+1} < \hat{t}_k +T/N$ for all $k < l^{*}$. It follows that (1) $\tau^{\max}_{k+1} = \hat{t}_{l^{*}} < \infty$ and (2) $\tau^{\min}_{k+2} = \hat{t}_{l^{*}+1} \geq \tau^{\max}_{k+1} + T/N$.

We conclude by showing that (3) holds. That is, 
\[
\sum_{n \geq 0} \textbf{1} \{ t^i_n \in [\tau^{\min}_{k+1} , \tau^{\max}_{k+1} ] \} \leq R+1, \quad \forall i \in \NN.
\]

By \Cref{assumption:exploration-phases}, it suffices to show that $\tau^{\max}_{k+1} - \tau^{\min}_{k+1} < (R+1)T .$

We have already shown $\hat{t}_0 - \tau^{\min}_{k+1} \leq RT - T/N$, which handles the case where $l^{*} = 0$, and we focus on $l^{*} > 0$. Since $\hat{t}_{k+1} < \hat{t}_k +T/N$ for all $k < l^{*}$ and $l^{*} \leq N-1$, we have $ \tau^{\max}_{k+1} - \tau^{\min}_{k+1}  = \hat{t}_{l^{*}} - \tau^{\min}_{k+1}  $ and 
\begin{align*}
\hat{t}_{l^{*}} - \tau^{\min}_{k+1}    = \left( \sum_{l = 0}^{l^{*} - 1} \left[ \hat{t}_{l + 1} - \hat{t}_{l} \right] \right) + \hat{t}_{0} -  \tau^{\min}_{k+1} 		\leq  l^{*} (T/N ) + ( RT - T/N  ) < (R+1)T,
\end{align*}
which concludes the proof. $ $
\end{proof}

Note that the proof above also shows that for any active phase $[ \tau^{\min}_k , \tau^{\max}_k ]$, there exists some agent that has exactly one opportunity to update its policy in $[ \tau^{\min}_k , \tau^{\max}_k ]$.

\

For $k \geq 0$, we define $B_k$ to be the event that the baseline policy is a fixed equilibrium policy throughout the $k^{th}$ active phase, $[\tau^{\min}_k, \tau^{\max}_k ]$. That is, 

\[
B_k := \left\{ \bphi_{\tau^{\min}_k} = \cdots = \bphi_{\tau^{\max}_k} \in \eq[0]_{SD} \right\}. 
\]

We define $L := \ell^{*} +1$, where $\ell^{*} = \max\{ \ell(\bpi) : \bpi \in \JointPolicySet_{SD} \}$ and $\ell ( \bpi )$ denotes the length of a shortest strict best-response path from $\bpi \in \JointPolicySet_{SD}$ to an equilibrium policy in $\eq[0]_{SD}$.  \label{def:L}

\vspace{2.5pt}

\begin{lemma}  \label{lemma:stay-at-equilibrium}
Let $\theta > 0$, and define 
\[
\xi := \frac{1}{(R+1) N L } \min \left\{ \theta, \frac{1}{2} \min_{i \in \NN} \{ \delta^i, \bar{\delta} - \delta^i \} \right\}.
\]

Suppose $\max_{i \in \NN} \alpha^i < \hat{\alpha}_{\xi}$ and $T \geq N \hat{T}_{\xi} ( \balpha )$, where $\hat{\alpha}_{\xi}$ and $\hat{T}_{\xi} ( \cdot )$ are the objects specified in \Cref{lemma:primitive-random-variables-lead-to-Q-convergence}. For any $k \geq 0$ and history $\bh_{\tau^{\max}_k} \in \bH_{\tau^{\max}_{k}, \rm eq}$, we have 
\[
\P \left( B_{k+1} | \bh_{\tau^{\max}_k} \right) \geq 1 - \theta/L. 
\]
\end{lemma}

\vspace{7.5pt}

\noindent \textbf{Remark:}  We note that $\tau^{\max}_k \in \zz_{\geq 0}$ is some constant, and $\bh_{\tau^{\max}_k}$ is a $\tau^{\max}_k$-history,
\[
\bh_{\tau^{\max}_k} = ( x_0, \bphi_0, \widehat{\bQ}_0, \dots, x_{\tau^{\max}_k}, \bphi_{\tau^{\max}_k}, \widehat{\bQ}_{\tau^{\max}_k} ),
\]
with $\bphi_{\tau^{\max}_k} \in \JointPolicySet_{SD} \cap \eq[0]$.

\vspace{7.5pt}

\begin{proof} We put $\tilde{t}_0 := \tau^{\min}_{k+1}$ and recursively define 
\[
\tilde{t}_{l+1} := \min \{ t^i_n > \tilde{t}_l : i \in \NN, n \geq 0 \}, \quad \forall l \geq 0.
\]

We define $m \in \zz_{\geq 0}$ to be the index achieving $\tilde{t}_m = \tau^{\max}_{k+1}$, and note that $m = 0$ is possible when $\tau^{\max}_{k+1} = \tau^{\min}_{k+1}$. 

Note that $m$ counts the number of stage games after $\tau^{\min}_{k+1}$ and on/before $\tau^{\max}_{k+1}$ at which \emph{any} player ends an exploration phase. From the proof of \Cref{lemma:active-phases} we have that 
\[
m \leq \sum_{i \in \NN} \sum_{n \geq 0} \textbf{1} \{ t^i_n \in [\tau^{\min}_{k+1} , \tau^{\max}_{k+1} ] \} <  (R+1)N   .  
\]

where the strict inequality obtains since at least one player (player $j(0)$ in the proof of \Cref{lemma:active-phases}) has exactly one exploration phase ending in $[\tau^{\min}_{k+1} , \tau^{\max}_{k+1} ]$, while the remaining $N-1$ players have at most $R+1$.

For each $l \in \{0, \cdots, m \}$, let $A_l$ denote the set of players who have an opportunity to switch policies at time $\tilde{t}_l$. That is,
\[
A_l := \{ i \in \NN : \exists \; n(l) \in \nn \text{ s.t. } t^i_{n(l)} = \tilde{t}_l \} . 
\]

From our choices of $\xi, \balpha$, and $T \geq N \hat{T}_{\xi} ( \balpha )$ and the fact that $\tilde{t}_l \geq \tau^{\min}_{k+1}  \geq \tau^{\max}_k + T/N$ for all $l \in \{ 0, 1, \dots, m \}$, we have, by \Cref{lemma:primitive-random-variables-lead-to-Q-convergence}, that
\[
\P \left( \Omega_{\tau^{\max}_k : \tilde{t}_l} \right) \geq 1 - \xi, \quad \forall l \in \{ 0, 1, \dots, m \},
\]

where, for any $s, t \geq 0$, we recall that $\Omega_{t: t+s}$ is given by
\[
\Omega_{  t: t+s}  = \left\{ \bvarpi_{t+} : \max_{i \in \NN} \left\| \bar{Q}^i_{t+s} ( \bh_{t}, \bvarpi_{t+} ) - Q^{*i}_{\bphi^{-i}_{t}} \right\|_{\infty} < \xi \right\} .
\]


Note that $\tilde{t}_0 = \tau^{\min}_{k+1}$, and the minimality defining $\tau^{\min}_{k+1}$ implies that no player updates its policy during the interval $(\tau^{\max}_{k}, \tau^{\min}_{k+1})$. It follows that the hypothetical continuation trajectory defining each player $i$'s hypothetical Q-factors $\bar{Q}^i_{\tilde{t}_0} ( \bh_{\tau^{\max}_k} , \bvarpi_{\tau^{\max}_k +} )$ coincides with the sample trajectory defining $\widehat{\bQ}_{\tilde{t}_0}$, since action selections and state transitions are decided by $\bphi_{\tau^{\max}_k}$ and $\bvarpi_{\tau^{\max}_k +}$. Thus, by our choice of $\tilde{t}_0 = \tau^{\min}_{k+1}$,
\[
\bar{Q}^i_{\tilde{t}_0} ( \bh_{\tau^{\max}_k} , \bvarpi_{\tau^{\max}_k +} )   = \widehat{Q}^i_{\tilde{t}_0} , \quad \forall i \in \NN . 
\]

For $\bvarpi_{\tau^{\max}_k +} \in \Omega_{ \tau^{\max}_k : \tilde{t}_0}$, each player $i \in A_0$ recovers its estimated Q-factors $\widehat{Q}^i_{\tilde{t}_0} $ within $\xi$ of  $Q^{*i}_{\bphi^{-i}_{\tau^{\max}_k}}$, since 
\[
\widehat{Q}^i_{\tilde{t}_0} = \QQ^i_{\tilde{t}_0} ( \bh_{\tau^{\max}_k}, \bvarpi_{\tau^{\max}_k +}  )   .
\]
Since we have hypothesized that $\bphi_{\tau^{\max}_k} \in \eq[0]_{SD}$ and chosen $\xi < \frac{1}{2}\min\{ \delta^i, \bar{\delta} - \delta^i \}$, it follows that agent $i$ does not change its policy at time $\tilde{t}_0 = \tau^{\min}_{k+1}$ when given the opportunity to revise its policy. In words, the agent appraises its policy as being sufficiently close to optimal with respect to its learned Q-factors, and does not switch its policy. In symbols, that is to say
\begin{align*}
			\bvarpi_{\tau^{\max}_k +} \in \Omega_{\tau^{\max}_k: \tilde{t}_0 }  \Rightarrow 	\max_{i \in \NN} \left\| \QQ^i_{\tilde{t}_0} ( \bh_{\tau^{\max}_k}, \bvarpi_{\tau^{\max}_k +} ) - Q^{*i}_{\bphi^{-i}_{\tau^{\max}_k}}  \right\|_{\infty} < \xi  ,
\end{align*}
which in turn implies that $\bphi_{\tilde{t}_0} = \Phi_{\tilde{t}_0} ( \bh_{\tau^{\max}_k}, \bvarpi_{\tau^{\max}_k +} ) = \bphi_{\tau^{\max}_k }. $

\

Repeating this logic, one has that if 
\[
\bvarpi_{\tau^{\max}_k +} \in \bigcap_{0 \leq l \leq m } \Omega_{\tau^{\max}_k : \tilde{t}_l }
\]
then $\bphi_{\tilde{t}_l} = \Phi ( \bh_{\tau^{\max}_k} , \bvarpi_{\tau^{\max}_k + } ) = \bphi_{\tau^{\max}_k}$ for each $l \leq m $.  The probability of this intersection is lower bounded using the union bound and \Cref{lemma:primitive-random-variables-lead-to-Q-convergence}:
\[
\P \left( \bigcap_{0 \leq l \leq m } \Omega_{\tau^{\max}_k : \tilde{t}_l } \right) \geq 1 - (m+1)\xi \geq 1 - \theta/L,
\]
as desired.  $ $ 
\end{proof}

\begin{lemma} \label{lemma:go-to-equilibrium}
Let $\theta > 0$, and define 
\[
\xi := \frac{1}{(R+1) N L } \min \left\{ \theta, \frac{1}{2} \min_{i \in \NN} \{ \delta^i, \bar{\delta} - \delta^i \} \right\}.
\]

Suppose $\max_{i \in \NN} \alpha^i < \hat{\alpha}_{\xi}$ and $T \geq N \hat{T}_{\xi} ( \balpha )$, where $\hat{\alpha}_{\xi}$ and $\hat{T}_{\xi} ( \cdot )$ are the objects specified in \Cref{lemma:primitive-random-variables-lead-to-Q-convergence}. For any $k \geq 0$ and history $\bh_{\tau^{\max}_k} \in \bH_{\tau^{\max}_{k}} \setminus \bH_{\tau^{\max}_{k}, \rm eq}$, we have 
\[
\P \left( B_{k+L} | \bh_{\tau^{\max}_k} \right) \geq  p_{\min} ( 1 - \theta ) ,
\]
where $p_{\min} := \prod_{j \in \NN} \min\left\{ \frac{1 - \lambda^j }{ | \IndividualPolicySet^j_{SD} |} , \lambda^j \right\}^{(R+1)L}  > 0$ and $L$ was defined above \Cref{lemma:stay-at-equilibrium}.
\end{lemma}

\vspace{5pt}

\begin{proof} Let $\bpi_0,\dots,\bpi_{\ell}$ be a (shortest) strict best-response path from $\bpi_0:=\bphi_{\tau_k^{\max}}$ to $\bpi_{\ell} \in  \eq[0]_{SD}$, and $\ell \in \{ 1 , \dots ,\ell^* \}$, where $\ell^{*}$ was defined along with $L$ immediately before \Cref{lemma:stay-at-equilibrium}. For each pair of neighbouring joint policies $\bpi_{l}$ and $\bpi_{l+1}$, there exists exactly one player $i(l)$ who switches its policy. That is, $\pi^j_{l+1} = \pi^j_l$ for all $j \not= i (l )$. 

Consider the event that no agent updates its policy during $[\tau_k^{\max},\tau_{k+1}^{\max}]$ due to inertia except player $i(0)$, who updates its policy exactly once to $\pi^{i(0)}_1$ and remains inert at all other update opportunities in $[\tau_k^{\max},\tau_{k+1}^{\max}]$.

By \Cref{lemma:primitive-random-variables-lead-to-Q-convergence} and \Cref{lemma:active-phases}, the conditional probability of this event given $\bh_{\tau_k^{\max}}$ is lower bounded by
\[
(1-\theta/L) \prod_{j \in \NN }    \min   \left\{     \frac{1-\lambda^j}{| \IndividualPolicySet_{SD}^j|},    \lambda^j  \right\}^{R+1} .
\]
The same lower bound similarly applies to each transition along $\bpi_0,\dots,\bpi_{\ell}$, which leads to
\begin{align*}
	\P\left( \bphi_{\tau^{\max}_{k + \ell}   }=\bpi_{\ell} \middle|  \bh_{\tau_k^{\max}}  \right) \geq (1-\theta/L)^{\ell} \prod_{ j \in \NN }   \min \left\{  \frac{1-\lambda^j}{|  \IndividualPolicySet_{SD}^j |},\lambda^j \right\}^{(R+1) \ell} .
\end{align*}

Next, consider the event that no agent updates its policy during $[\tau^{\max}_{k+\ell}, \tau^{\max}_{k+ L} ]$ due to inertia.
The conditional probability of this event given $h_{\tau^{\max}_{k + \ell}}$ is lower bounded by\
\[
\prod_{  j \in \NN }   (\lambda^j)^{(R+1)(L-\ell)   } .
\]

This results in $ \P(B_{k+L} |  \bh_{\tau_k^{\max}}) \geq (1-\theta/L)^L p_{\min}, $ which proves the lemma since $(1-\theta/L)^L  \geq 1-\theta$. 
\end{proof}

\subsection{Proof of \Cref{theorem:main}}

Given $\epsilon>0$, let $\theta>0$ be the unique solution to 
\[
\frac{(1-\theta)   p_{\min}}{\theta + (1-\theta)  p_{\min}}-\theta = 1 -\epsilon
\]
and
\[
\xi=\frac{1}{(R+1)NL}\min \left\{\theta,\frac{1}{2}\min_{i \in \NN}\{\delta^i,\bar{\delta}-\delta^i\} \right\}.
\]
Suppose that 
\[
\max_{ i \in \NN } \alpha^i \leq \bar{\alpha}_{\epsilon} := \hat{\alpha}_{\xi} \: \, \textrm{and} \quad T \geq \bar{T}_{\epsilon}(\balpha,R):=N\hat{T}_{\xi}(\balpha).
\]

By Lemmas~\ref{lemma:stay-at-equilibrium} and \ref{lemma:go-to-equilibrium}, for all $k \geq 0$,
\begin{align}
\P(B_{k+L}|B_k) & \geq 1-\theta 			,	\label{eq:stay-at-equilibrium}\\
\P(B_{k+L}|B_k^c) & \geq (1-\theta) p_{\min} 	.	\label{eq:go-to-equilibrium}
\end{align}

Let $p_k:=\P(B_k)$ for all $k \geq 0$. The subsequent details lower bounding $p_{k+mL}$ for large $m$ and every $k < L$ are omitted, as they resemble the proofs of \cite{arslan2017decentralized} or \cite{yongacoglu2023satisficing}.%

\vspace{20pt}

\textbf{Remark:} In the interest of clarity, \Cref{theorem:main} was stated and proved for a special case of deterministic exploration phase lengths $\{ T^i_k : i \in \NN, k \geq 0\}$. The analysis above can be immediately extended to handle random exploration phase lengths, provided these are primitive random variables (as opposed to being history dependent) and provided they satisfy the requisite assumptions appearing in \Cref{theorem:main} almost surely.

\section{Simulations}  \label{sec:simulations}

In this section, we present empirical findings of a simulation study where Algorithm~\ref{algo:main} was applied to the stochastic game whose stage game costs and transition probabilities are presented below.

We have chosen to simulate play of a game with two players, $\NN = \{ 1,2 \}$, state space $\xx = \{ s_0, s_1 \}$, and two actions per player, with action sets labelled $\IndividualActionSet^1 = \IndividualActionSet^2 = \{ a_0, a_1 \}$. Each player's discount factor is set to 0.8, and the transition kernel is described as follows: 
\begin{align*}
\P \left( x_{t+1} = s_0 | x_t = s_0 , a^1_t ,  a^2_t   \right) 	&\equiv 0.5 , \\ 
\P \left( x_{t+1} = s_0 | x_t = s_1 , a^1_t = a^2_t \right) 					&= 0.25 			\\
\P \left( x_{t+1} = s_0 | x_t = s_1 , a^1_t \not= a^2_t \right) 				&= 0.9 										
\end{align*}

In words, the state dynamics out of state $s_0$ do not depend on the choice of joint action, while the dynamics out of state $s_1$ do. In state $s_1$, if players select matching actions, then the state transitions to $s_0$ with (low) probability 0.25. Otherwise, if players select distinct actions in state $s_1$, play transitions to state $s_0$ with (high) probability 0.9.

The stage game's symmetric cost structure is summarized by the tables below. 

\begin{figure}[h]
\centering
\begin{subfigure}{.5\textwidth}
  \centering
\begin{game}{2}{2}[]
			&$a_0$      		&$a_1$					\\
$a_0$ 		&0, 0	  			&2, 2 						\\
$a_1$		&2, 2				&0, 0				
\end{game}
  \caption{State $s_0$: low cost state}
  \label{fig:sub1}
\end{subfigure}%
\begin{subfigure}{.5\textwidth}
  \centering
\begin{game}{2}{2}[]
			&$a_0$      					&$a_1$					\\
$a_0$ 		&10, 10	  					&11, 11					\\
$a_1$		&11, 11						&10, 10	
\end{game}
  \caption{State $s_1$: high cost state}
  \label{fig:sub2}
\end{subfigure}
\caption{Stage costs: Player 1 (respectively, 2) picks a row (column), and its stage cost is the 1st (2nd) entry of the chosen cell.}
\label{fig:two-player-game}
\end{figure}


In either state, the stage cost to a given player is minimized by selecting one's action to match the action choice of the opponent. We note that the stage costs in state $s_0$ are considerably lower than those of state $s_1$. Despite the apparent coordination flavour of the stage game payoffs, the transition probabilities dictate that an optimizing agent should \emph{not} match its counterpart's action in state $s_1$, and should instead attempt to drive play to the low cost state $s_0$. 

The challenge of non-stationarity to multi-agent learning in this game is such: if players use a joint policy under which they mismatch actions in state $s_0$ or match actions in state $s_1$, then either player has an incentive to switch its policy in the relevant state. If both players switch policies in an attempt to respond to the other, then cyclic behaviour ensues, where the individual responds to an outdated policy, as the other player will have switched its behaviour by the time the individual is ready to update its own policy.  

With the cost structure, transition probabilities, and discount factors described above, we have the joint policies in which players' actions match in state $s_0$ and players' actions do not match in state $s_1$ constitute 0-equilibrium policies.

\subsection*{Parameter Choices}

We conducted 500 independent trials of Algorithm~\ref{algo:main}. Each trial consisted of $10^6$ stage games. Each player's exploration phase lengths, $T^i_k$, were selected uniformly at random from the interval $[ T, R \cdot T ]$, 
with $T$ set to $5000$ and the ratio parameter $R$ set to $R=3$.

For our gameplay parameters, we set the stage game action experimentation parameters to be $\rho^1 = \rho^2 = 0.05$. The parameter controlling inertia in the policy update was chosen as $\lambda^1 = \lambda^2 = 0.2$. The parameter controlling tolerance for sub-optimality in Q-factor assessment was chosen as $\delta^1 = \delta^2 = 0.5$. The learning rate parameters were selected to be $\alpha^1 = \alpha^2 = 0.08$. 

\subsection*{Results} The results of our simulations are summarized below in Figure~\ref{fig:graph} and Table~\ref{table:some-results}. With randomly selected initial policies, we observed that Algorithm~\ref{algo:main} rapidly drove play to equilibrium, with the relative frequency of equilibrium stabilizing at over 98\% of trials after approximately 40,000 stage games. 

\begin{figure}[h] 
\includegraphics[width = 0.85\textwidth]{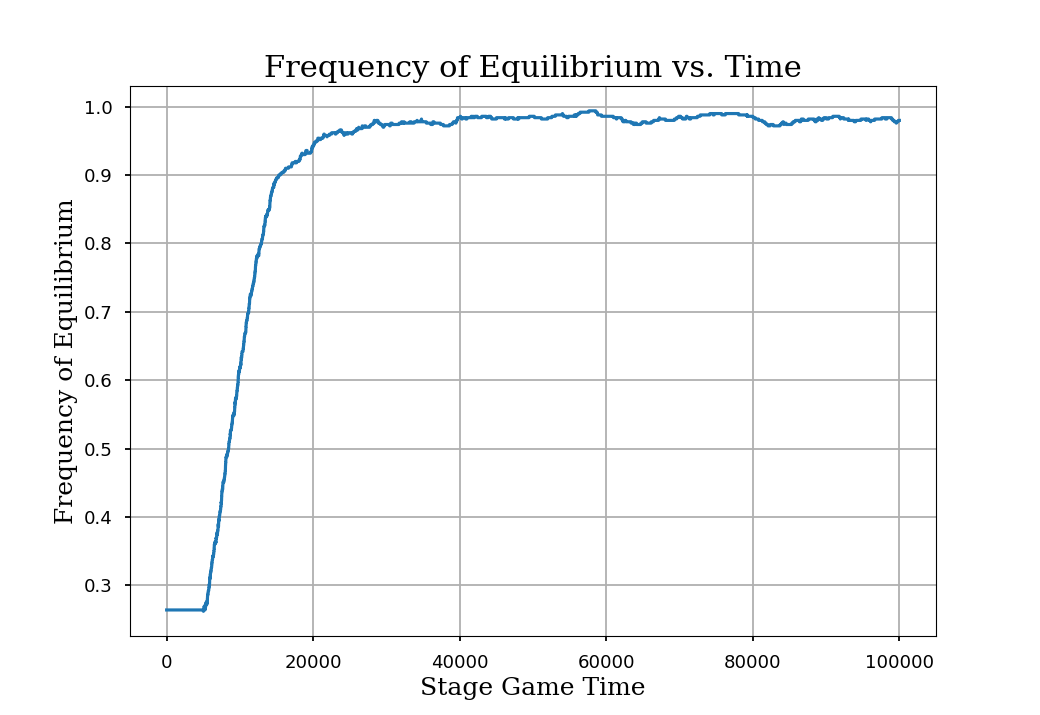}   %
\centering
\caption[]{Frequency of $\bphi_t \in \eq[0]_{SD}$, averaged over 500 trials.}
\label{fig:graph}
\end{figure}

\begin{table}[h] \centering
\renewcommand{\arraystretch}{1.5}
\begin{tabular}{ c | c | c | c | c | c  }
Stage game index $t =$										& 0 		& 10,000		& 20,000		&30,000 		& 40,000 	\\ 
\hline
$\frac{1}{500} \sum_{k=1}^{500} \textbf{1} \{ \bphi_t \in \eq[0]_{SD} \}$	&0.264	& 0.620		& 0.728		&0.974		&0.984		
\end{tabular}
\caption{Frequency of equilibrium at select stage game indices}   \label{table:some-results}
\end{table}

\section{Conclusions}

In this paper, we considered an unsynchronized variant of the decentralized Q-learning algorithm of \cite{arslan2017decentralized}. We formalized an argument presented in \cite{Marden2009payoff}, which contended that inertia alone would stabilize the learning process and tame non-stationarity. In the process of formalizing the argument of \cite{Marden2009payoff}, some modifications were needed to handle technical nuances relating to the dynamics of the policy process and the convergence of learning iterates. To accommodate unsynchronized policy updating and  non-stationarity in each agent's learning environment, we have introduced a constant learning rate that can rapidly overcome errors in learning estimates that are artifacts of outdated information. In so doing, we have shown that decentralized Q-learning can still drive policies to equilibrium in weakly acyclic stochastic games without making strong coordination assumptions such as synchronizing the schedules on which players update their policies.

\bibliographystyle{plain}
\bibliography{unsynchronized_2024}

\newpage

\appendix

\color{black}


\newpage
\section{Glossary of Notation}   \label{appendix:notation}

\begin{table}[h]
\color{black}
\begin{center}
\renewcommand{\arraystretch}{1.25} 
\begin{tabular}{|l|p{0.5\linewidth}| c | }
\hline
\textbf{Notation} 		& \textbf{Description} 			& \textbf{Reference(s)} \\
\hline
$Q^{*i}_{\bpi^{-i}}$		&  Q-function when facing $\bpi^{-i} \in \JointPolicySet^{-i}_{S}$    & \eqref{eq:Q-factors} \\[2pt]
\hline
$\widehat{Q}^i_t$		& Player $i$'s Q-factor estimates at time $t$	& 	\eqref{eq:single-agent-Q-factor-update}, Algorithm~\ref{algo:main} \\
\hline
$\widehat{\bQ}_t$		&  $\left ( \widehat{Q}^1_t, \dots, \widehat{Q}^N_t \right) $		& p. \pageref{ss:more-notation}   \\ 
\hline
EP					& Exploration Phase				& \\
\hline
$T^i_k$ 				& Length of player $i$'s $k^{th}$ EP 	&Algorithm~\ref{algo:main}  \\
\hline
$t^i_k$				& Start time of  player $i$'s  $k^{th}$ EP 	& Algorithm~\ref{algo:main}  \\
\hline
$\rho^i \in (0,1)$		& Player $i$'s action experimentation parameter 	&  	Algorithm~\ref{algo:main} \\
\hline
$\lambda^i \in (0,1)$ 		& Player $i$'s policy inertia parameter      & 	Algorithm~\ref{algo:main}  \\
\hline
$\delta^i > 0 $ 			& Player $i$'s  ``tolerance for suboptimality" 		& 	Algorithm~\ref{algo:main} \\
\hline
$\alpha^i \in ( 0, 1 ) $ 	& Player $i$'s learning rate/step size 	& 	Algorithm~\ref{algo:main} \\
\hline
$\balpha$				& Tuple of learning rates: $(\alpha^i )_{i \in \NN}$			& p. \pageref{assumption:exploration-phases}  \\ 
\hline
$\pi^i_k $ 				& Player $i$'s baseline policy during its $k^{th}$ EP 		& Algorithm~\ref{algo:main} \\
\hline
$\phi^i_t$				&Player $i$'s baseline policy for stage game at time $t$	& p. \pageref{assumption:exploration-phases}  \\
\hline
\end{tabular}
\end{center}
\caption{Algorithm-specific Notation}  \label{table:algo-notation}
\end{table}

\begin{table}[h]
\color{black}

\begin{center}
\renewcommand{\arraystretch}{1.25} 
\begin{tabular}{|l|p{0.5\linewidth}| c | }
\hline
\textbf{Notation} 		& \textbf{Description} 				& \textbf{Reference(s)} \\
\hline
$\bar{\rho}$			& An upper bound on each player's experimentation parameters $\rho^i$		& \Cref{assumption:rho-delta}    \\ 
\hline
$\bar{\delta}$			& An upper bound on each player's  $\delta^i$				&  \Cref{assumption:rho-delta} \\ 
\hline
$\hat{\pi}^i_k$			& Player $i$'s behaviour policy during $k^{th}$ EP: $\hat{\pi}^i_k$ selects according to $\pi^i_k$ w.p. $(1-\rho^i)$ and selects from $\uniform( \IndividualActionSet^i)$ w.p. $\rho^i$.			& p. \pageref{eq:hat-policy} \\ 
\hline		
$T$					& Lower bound on EP lengths: $T \leq T^i_k$		 			&  \Cref{assumption:exploration-phases}  \\ 
\hline
$R$					& Factor bounding EP lengths: $T^i_k \leq RT$						& \Cref{assumption:exploration-phases}  \\ 
%
\hline
$[ \tau^{\min}_k ,\tau^{\max}_k]$	& Active phase of policy updating   & \Cref{def:active-phases} \\
\hline 
$B_k$ 		& Equilibrium event defined using $k^{th}$ active phase $[ \tau^{\min}_k ,\tau^{\max}_k]$	& p. \pageref{lemma:stay-at-equilibrium} 		\\
\hline 
\end{tabular}
\end{center}
\caption{Analysis-specific Notation (Part 1)}  \label{table:analysis-notation}
\end{table}

\newpage

\begin{table}[h]
\color{black}

\begin{center}
\renewcommand{\arraystretch}{1.25} 
\begin{tabular}{|l|p{0.5\linewidth}| c | }
\hline
\textbf{Notation} 		& \textbf{Description} 				& \textbf{Ref(s)} \\
\hline
$\{ W_t \}_{t \geq 0}$		& An i.i.d. noise process taking values in $[0,1]$. 			&p. \pageref{ss:primitive-random-variables} \\ 
\hline
$f{\!}:{\!}\xx{\!}\times{\!}\JointActionSet{\!}\times [0,1]{\!}\to{\!}\xx$	& A function for expressing state transitions as a deterministic function driven by exogenous noise: $x_{t+1} = f( x_t, \ba_t, W_t )$			&p. \pageref{ss:primitive-random-variables} \\ 
\hline
$ \{ \tilde{u}^i_t  \}_{t \geq 0}$ 	& An i.i.d. sequence of actions for player $i$, with $\tilde{u}^i_t \sim \uniform( \IndividualActionSet^i )$						&p. \pageref{ss:primitive-random-variables} \\ 
\hline
$\tilde{\rho}^i_t$		& An i.i.d. sequence   with $\tilde{\rho}^i_t \sim \uniform( [0,1])$ 				&p. \pageref{ss:primitive-random-variables} \\ 
\hline
$\tilde{\lambda}^i_t$		& An i.i.d. sequence   with $\tilde{\lambda}^i_t \sim \uniform( [0,1])$  						&p. \pageref{ss:primitive-random-variables} \\ 
\hline
$\{ \tilde{\pi}^i_t (B^i ) \} $		&An i.i.d. sequence of policies for player $i$, with  $\tilde{\pi}^i_t ( B^i ) \sim \uniform( B^i )$, where $B^i \subseteq \IndividualPolicySet^i_{SD}$,  	&p. \pageref{ss:primitive-random-variables} \\
\hline
$\omega^i_t$			& Random variables for player $i$ at time $t$:  $\omega^i_t = ( \tilde{\rho}^i_t, \tilde{u}^i_t, \tilde{\lambda}^i_t, \tilde{\pi}^i_t ( B^i )  : B^i \subseteq \IndividualPolicySet^i_{SD}  ) $			& p. \pageref{ss:more-notation}  \\ 
\hline
$\mathfrak{S}^i$		& Codomain of $\omega^i_t$: $\omega^i_t \in \mathfrak{S}^i$		& p. \pageref{ss:more-notation}  \\ 
\hline
$\bomega_t$			& Collected state noise and player noise for time $t$: $ \bomega_t = (W_t, \omega^1_t, \dots, \omega^N_t)$ 			& p. \pageref{ss:more-notation}   \\ 
\hline
$\bvarpi_{t +}$	& Present and future of the noise processes: $ \bvarpi_t = (\bomega_t, \bomega_{t+1}, \dots)$	& 	p. \pageref{ss:more-notation}   \\
\hline
$\bh_t$				& History of states, policies, and Q-iterates to time $t$: $(x_0, \bphi_0, \widehat{\bQ}_0 , \dots, x_t, \bphi_t, \widehat{\bQ}_t)	$	& 	p. \pageref{ss:more-notation}   \\
\hline
$\bH_t$				& Codomain of $\bh_t$: $\bh_t \in \bH_t$		& 		p. \pageref{ss:more-notation}   \\
\hline
$\bH_{t, {\rm eq}} \subset \bH_t$			& $\{ \bh_t \in \bH_t : \bphi_t \in \eq[0]_{SD}   \} $		& 		p. \pageref{ss:more-notation}   \\
\hline
$\QQ^i_t$			&Mapping that reports Q-iterates at time $t$ using historical data $\bh_s$ up to time $s$ and future random variables from time $s$ onward in $\bvarpi_s$:   $\QQ^i_t ( \bh_s , \bvarpi_s  )  = \widehat{Q}^i_t $ for all $0 \leq s \leq t$. 		& \eqref{def:Q-and-Phi-mappings}    \\
\hline
$\Phi_t$			& Mapping that reports baseline policies at time $t$ using historical data to time $s$ and future random variables from $s$ onward : $\Phi_t ( \bh_s, \bvarpi_s ) = \bphi_t $ for all $0 \leq s \leq t$.  		& \eqref{def:Q-and-Phi-mappings}    \\
\hline
$\bar{u}^i_{t+m} ( \bh_t, \bvarpi_t )$		& Action chosen by player $i$ at time $t+m$ in hypothetical scenario where (i) baseline policies were frozen at $\bphi_t$ at time $t$, and (ii) action selection and state transition noise was described by $\bvarpi_t$	& p. \pageref{ss:supporting-results}    \\
\hline
$\bar{x}_{t+m} ( \bh_t, \bvarpi_t ) $			& State at time $t+m$ in hypothetical scenario where (i)   and (ii)  above hold.	& p. \pageref{ss:supporting-results}    \\
\hline 
 $\bar{Q}^i_{t+s} ( \bh_t, \bvarpi_t )$	& Hypothetical Q-factors that would have been obtained at time $t+s$  in hypothetical scenario where (i)   and (ii)  above hold.	& p. \pageref{ss:supporting-results}    \\    
\hline
\end{tabular}
\end{center}
\caption{Analysis-specific Notation (Part 2)} \label{table:analysis-notation-continued}
\end{table}

\end{document}